\documentclass[a4paper,11pt]{article}
\pdfoutput=1 % if your are submitting a pdflatex (\ie if you have
\usepackage{pub} 
\usepackage[T1]{fontenc} % if needed
\usepackage{subfigure}
\usepackage{amssymb}
\usepackage{abbreviations}
\usepackage{mathrsfs}
\usepackage{amsthm}
\usepackage{xspace}
\usepackage{lineno}
\usepackage{multirow}

\graphicspath{{./fig/}}

\title{\boldmath Performance and optimization of support vector machines in high-energy physics classification problems}
 \author[1]{M.\"O.~Sahin\note{Corresponding author.}}
 \author{D.~Kr\"ucker}
 \author{and I.-A.~Melzer-Pellmann}
 \affiliation{Deutsches Elektronen-Synchrotron,\\Notkestr.\ 85, 22607 Hamburg, Germany}
\emailAdd{ozgur.sahin@desy.de}
\emailAdd{dirk.kruecker@desy.de}
\emailAdd{isabell.melzer@desy.de}
\abstract{
In this paper we promote the use of Support Vector Machines (SVM) as a machine learning tool for searches in high-energy physics.
As an example for a new-physics search we discuss the popular case of Supersymmetry at the Large Hadron Collider. 
We demonstrate that the SVM is a valuable tool and show that an automated discovery-significance based optimization of the SVM  hyper-parameters is a highly efficient way to prepare an SVM for such applications. A new C++ LIBSVM interface called SVM-HINT is developed and available on Github. 
}
\begin{document} 
\maketitle
\flushbottom
%\linenumbers

% sections
% !TEX root = svm_paper.tex

\section{Introduction}
\label{sec:intro}

Data analysis in High Energy Physics (HEP) is a genuine multivariate problem. Despite the fact that multivariate 
techniques have been used in HEP for a long time, the explosive growth of machine learning (ML) techniques during the 
last two decades had only a limited impact on the accustomed style in which data analysis is performed in this field. 
TMVA~\cite{Hocker:2007ht} is probably the most commonly used software package in this context and especially 
Boosted Decision Trees (BDT) and Artificial Neural Networks are applied to explore the large and complex datasets 
delivered by present-day experiments. Only recently an increased interest in the machine learning expertise acquired 
in other areas of science can be observed~\cite{higgsml,higgsmlProc,dataSienceLHC15}.

In this paper we promote the application of Support Vector Machines (SVM)~\cite{Boser:1992,CorVap95,vapnik98a}
for new-physics searches. Support Vector Machines are a competitive and widely used approach to binary 
classification. The search for new physics can be considered as a classification task where the rare new-physics 
signal and the dominant Standard Model (SM) background constitute the two distinct classes. Although there are a few HEP 
papers on SVMs~\cite{Vannerem:1999wm,Vaiciulis:2002jw,Janyst2008,PhysRevD.86.032011,Sforza2011}, this approach seems 
to be heavily undervalued amongst HEP researchers considering the many thousands of publications on SVM applications 
that a simple literature search yields.

After an introduction to SVMs in Section~\ref{SVM}, the hyper-parameter tuning is described in Section~\ref{sec:tuning}
(and Appendix~\ref{app:itergrid}) within the context of our new SVM framework: {\sc SVM-Hint}. 
In Section~\ref{sec:cases} we first discuss a toy 
model and then an example for an actual new-physics search, targeted at a supersymmetric partner of the top quark 
at the LHC. We demonstrate that the SVM is a valuable tool for HEP searches and that the partial neglect of the 
SVM approach within the HEP community can be related to the limited performance of its implementation in TMVA without
an automated hyper-parameter search. Moreover, we show that a significance-based optimization of the 
hyper-parameters is a highly efficient way to prepare an SVM for a HEP search. 

In addition, we provide a software package~\cite{ourgithub} that performs such a significance-based optimization of hyper-parameters 
and interfaces ROOT~\cite{Antcheva:2011zz} trees with the popular SVM implementation LIBSVM~\cite{CC01a}.

% !TEX root = svm_paper.tex
% !TEX spellcheck = en-US
\section{Support vector machines}\label{SVM}

A HEP search typically starts with a set of physical variables and cuts on these variables. The cuts are defined to 
select a new-physics signal against the background of known physics and are often chosen in a more or less ad-hoc style. 
The optimal use of these variables is a typical machine learning problem.
Monte Carlo (MC) simulation samples for the signal and background class can be used to train a 
supervised\footnote{A ML algorithm is called \textsl{supervised} when the class membership of all training vectors is known.}
machine learning algorithm which is potentially a more efficient classifier than a set of cuts.

We write for a set of $N$ training events: 
\begin{align}
\label{for:tra}
&(y_1,\mb{x}_1),(y_2,\mb{x}_2),...,(\mb{x}_i,y_i),...,(y_N,\mb{x}_N)\qquad y_i \in \{-1,1\},\\
&\mb{x}_i=(x_i^{(1)},\dots,x_i^{(n)})\label{for:tra2}
\end{align}
where for an event $i=1 \dots N$  the label $y_i$ distinguishes between signal and background and  $\mb{x}_i$ is an $n$-dimensional 
vector formed from the physical variables under consideration. These vectors constitute an n-dimensional real vector space $\mathbb{V}$.

A support vector machine is a supervised binary classifier based on the intuitive concept of an n-dimensional hyperplane separating 
two distinct classes. In this approach, finding the best separating hyperplane is considered to be a convex optimization problem. 
In its simplest form a SVM defines the eponymous \textsl{support vectors} as those elements of the training sample which are closest 
to the hyperplane. The separation margin between the classes is completely defined by the support vectors and maximized by the algorithm. 
%Using the plane with the maximum margin is an application of the principle of structural risk minimization\cite{vapnik77}.
This idea can be extended to overlapping distributions and eventually, by an implicit transformation of the variables, known as the 
\textsl{Kernel trick}, to non-linear problems. The last two modifications introduce additional hyper-parameters that must be set to some 
best value before the SVM training. In the following we give a short introduction to the concepts behind the SVM algorithm and to the
hyper-parameter tuning. The reader who is more interested in applications may continue with Section~\ref{sec:cases}. 

\subsection{Linearly separable distributions}
\begin{figure}[t] 
\centering
\includegraphics[width=0.6\textwidth]{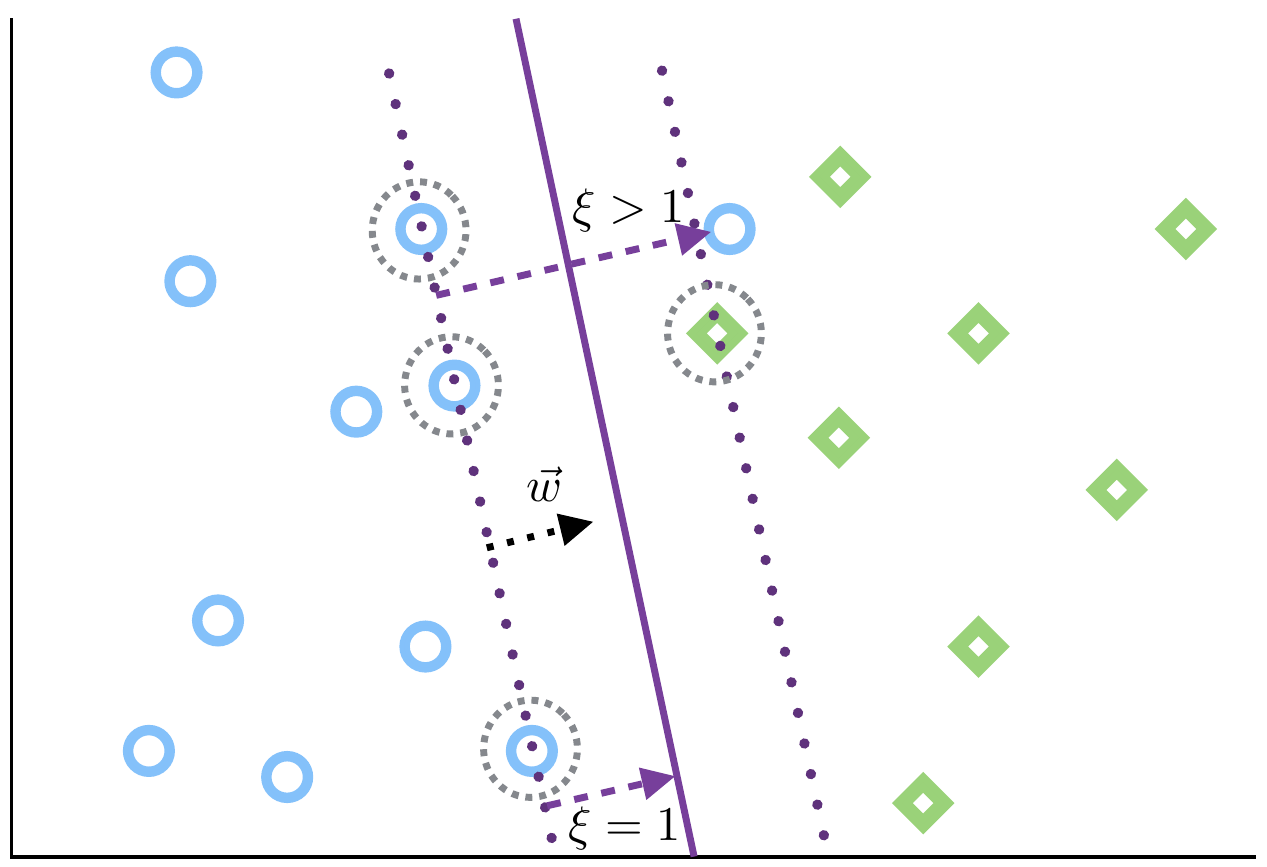}
\caption{The events represented by blue circles belong to the first class ($y=-1$), whereas the green triangles belong to the second 
class ($y=1$). The dashed lines represent the maximum margin boundaries, and the corresponding support vectors are circled by dashed 
lines. From all possible hyperplanes dividing the two samples, the one with the largest margin is chosen. The blue circle at $\xi>1$ 
is not linearly separable, see Sec.~\ref{sec:over}.
}
\label{fig:svm}
\end{figure}
A linear SVM separates the elements of two classes by an optimal hyperplane. Optimal in this approach is a hyperplane that maximizes 
the margin between the two classes for a given training sample. Those elements of the training sample sitting on the maximum margin 
boundaries are called support vectors. The support vectors are sufficient to construct the optimal hyperplane. 
Fig.~\ref{fig:svm} illustrates these ideas.

A separating hyperplane can be described as  $\mb{w}\!\cdot\!\mb{x}+b=0$, $\mb{w} \in \mathbb{V}$ and $b \in \mathbb{R}$. The vectors 
of the training sample are either \textsl{above} or \textsl{below}\footnote{For simplicity we use a 3 dimensional way of speaking. 
All described concepts are valid in $n$ dimensions. } this plane. We can always choose the scale of $\mb{w}$ and $b$ such that for the 
vectors which are closest to the hyperplane, \ie the support vectors $\mb{x}_k$, we obtain $\mb{w}\!\cdot\!\mb{x}_k+b=\pm 1$. Multiplied 
with the class label $y_i$, this expression must always be positive for correctly classified points:
\begin{equation}
\label{for:constraint}
y_{i}(\mb{w}\!\cdot\!\mb{x}_i+b)-1 \geqslant 0\;,
\end{equation}
and the equality is satisfied by the support vectors (circled in Fig.~\ref{fig:svm}). 
The separation margin is the distance $\rho$ between the support vectors on both sides. With the normal vector to the hyperplane 
$\mb{w}/|\mb{w}|$ and two arbitrary support vectors from each side $\mb{x}_{k+,}\mb{x}_{k-}$ the margin is given by:
\begin{equation}\label{for:mini1}
\rho(\mb{w},b) =  \frac{\mb{w}\!\cdot\!\mb{x}_{k+}}{|\mb{w}|} - \frac{\mb{w}\!\cdot\!\mb{x}_{k-}}{|\mb{w}|} = \frac{2}{|\mb{w}|}\;, 
\end{equation}
where the second equality follows from Eq.~\ref{for:constraint}. Maximizing the margin $\rho=2/|\mb{w}|$ is equivalent to minimizing $|\mb{w}|^2$. 
Finding the optimal separating hyperplane is therefore identical to solving the following quadratic optimization problem where the correct 
classification is enforced by the constraints from Eq.~\ref{for:constraint}:
\begin{equation}\begin{array}{rl}
\min  \limits_{\rule{0pt}{1em} \mb{w} \in \mathbb{V}, \; b \in \mathbb{R} }  \;\; &\frac{1}{2}\,|\mb{w}|^2\\
\mbox{subject to}\;\; &y_{i}(\mb{w}\!\cdot\!\mb{x}_i+b) \geqslant 1 \;\; \mbox{for all}\; i=1 \dots N\;.
\end{array}\end{equation}
This is a quadratic optimization problem with inequality constraints and can be solved using 
Lagrange multipliers $\alpha_i$ (with $\alpha_i \geqslant 0$). The Lagrangian can be written as:
\begin{equation}
\label{for:lagrange}
\mathscr{L}=\frac{1}{2}|\mb{w}|^2-\sum_{i=1}^N \alpha_i [y_{i}(\mb{w}\!\cdot\!\mb{x}_i+b)-1]\;.
\end{equation}
The solution is a saddle point $(\mb{w}^0,b^0,\alpha_i^0)$ where the Lagrangian becomes minimal with respect to $\mb{w}$ and $b$ and where 
the derivatives are:
\begin{eqnarray}
\label{for:w}
\frac{\partial\mathscr{L}}{\partial\mb{w}} =0&=&\mb{w}-\sum_{i=1}^N\alpha_i y_i \mb{x}_i\;, \\
\frac{\partial\mathscr{L}}{\partial b} =0&=&\sum_{i=1}^N \alpha_i y_i\;.
\label{for:w2}
\end{eqnarray}
Substituting these conditions into Eq.~\ref{for:lagrange} results in the dual Lagrangian: 
\begin{equation}
\label{for:lagrangefin}
\mathscr{L}=-\frac{1}{2}\sum \limits_{i=1}^N \sum \limits_{j=1}^N \alpha_i \alpha_j y_i y_j \, \mb{x}_i \!\cdot\! \mb{x}_j +\sum_{i=1}^N \alpha_i \;.
\end{equation}
Most SVM implementations search for a numerical solution to this dual problem. 
The dual Lagrangian is maximized with respect to $\alpha_i$ and must fulfill
the Karush-Kuhn-Tucker (KKT) conditions~\cite{karush,kuhntucker}:
\begin{equation}\label{for:kkt2}
\alpha_i\geqslant 0, \qquad \alpha_i(y_i(\mb{w}\!\cdot\!\mb{x}_i +b)-1)=0; \qquad i=1\dots N.
\end{equation}
Due to these conditions together with Eq.~\ref{for:constraint}, all non-support vectors are forced to have vanishing 
Lagrange multipliers $\alpha_i=0$, and only the support vectors contribute to the sums in Eq.~\ref{for:w} and \ref{for:w2}, 
and in the calculation of $\mb{w}_0$ and $b_0$ at the optimum.

The decision function, \ie the expression to predict the class label $\hat{y}$ of a new vector $\mb{u}$, follows 
from the hyperplane equation at the optimum:
\begin{equation}\label{for:dec}
\hat{y}=\sign(\sum_{k=1}^{N_{SV}} y_k \alpha_k \mb{x}_k\!\cdot\! \mb{u}-b_0)\;,
\end{equation}
where $N_{SV}$ is the number of support vectors.

It is important to note that the dual form in Eq.~\ref{for:lagrangefin} only depends on the scalar products of input vectors, 
and the same is true for the decision function Eq.~\ref{for:dec}. This advantage of the dual form is essential for 
the non-linear case in Sec.~\ref{sec:nonlin}.

\subsection{Overlapping distributions} \label{sec:over}
The method described so far works in the case of linearly separable data. Overlapping signal and background distributions 
require a different treatment. By allowing misclassification, the \textsl{hard margin} separation above can be modified 
into a \textsl{soft margin} approach. This can be done by introducing \textsl{slack} variables~\cite{CorVap95} 
($\xi_i\geqslant 0,\;i=1\dots N$) which measure for each training vector the relative distance by which they are on the 
wrong side of the separating hyperplane (shown for the one misclassified point in Fig.~\ref{fig:svm}). They are used to 
weaken the constraints~\ref{for:constraint}:  
\begin{equation}
\label{for:softconst}
y_{i}(\mb{w}\!\cdot\!\mb{x}_i+b) \geqslant 1-\xi_i\;,
\end{equation}
and allow to introduce the sum of the slacks $\sum_i^N\!\xi_i$ as a penalty term into the optimization problem.
The modified Lagrangian becomes:
\begin{equation}\label{for:lag2}
\mathscr{L}=\frac{1}{2}|\mb{w}|^2+C\sum \limits_i \xi_i-\sum \alpha_i [y_{i}(\mb{w}\!\cdot\!\mb{x}+b)-1+\xi_i]-\sum \beta_i \xi_i\;,
\end{equation}
and the extremum condition $\partial \mathscr{L}/\partial \xi_i=0$ implies a relation between the Lagrange multipliers, 
$\beta_i = C - \alpha_i$, which allows to bring \ref{for:lag2} into the same form as \ref{for:lagrange} and eventually 
into the dual form \ref{for:lagrangefin}. We are left with an optimization problem that is identical to the hard margin 
case up to the modified constraints.  

The constant $C$ that controls the strength of the penalty term appears now only as an upper limit on the Lagrange 
multipliers $0\leqslant\alpha_i\leqslant C$, restoring the hard margin case in the limit of $C\rightarrow \infty$.  
Furthermore, it controls the trade-off between simplicity of the decision rule and error frequency and is one of 
the hyper-parameters that must be set to a sensible value \textsl{before} the SVM training. 

\subsection{Non-linear distributions}\label{sec:nonlin}
The linear SVM presented in the two previous sections is quite limited. For HEP searches we expect complicated, non-linear hyper-surfaces
separating the two classes, for which the presented approach can easily be extended to create non-linear decision boundaries. 
The basic idea for a non-linear SVM~\cite{Boser:1992} is to map the input vectors $\mb{x}_i$ into a higher dimensional \textsl{feature space} 
$F$ where the problem becomes linearly separable: $\mb{x}_i \mapsto \mb{\phi}(\mb{x}_i) \in F$. The construction of a linear SVM in 
this feature space follows the same lines as before and the dual Lagrangian from Eq.~\ref{for:lagrangefin} will contain an inner 
product $\langle \mb{\phi}(\mb{x}_i), \mb{\phi}(\mb{x}_j) \rangle$ of elements of $F$.
The peculiar fact that the input vectors only appear in the dual Lagrangian, as well as in the decision function of Eq.~\ref{for:dec}, 
in form of scalar products allows us to avoid the explicit mapping and to use instead a kernel function  
${\mathrm K}(\mb{x}_i,\mb{x}_j) \equiv \langle \mb{\phi}(\mb{x}_i), \mb{\phi}(\mb{x}_j) \rangle $,
such that the dual Lagrangian and the decision function become 
\begin{eqnarray}
\mathscr{L}&=&-\frac{1}{2}\sum \limits_i \sum \limits_j \alpha_i \alpha_j y_i y_j \,{\mathrm K}(\mb{x}_i , \mb{x}_j)+\sum \alpha_i,\\
\hat{y}&=&\sign(\hat{f}),\qquad\hat{f}=\sum_{k=1}^{N_{SV}} y_k \alpha_k \, {\mathrm K}(\mb{x}_k,\mb{u})-b_0. \label{for:deci}
\end{eqnarray}
The existence of the mapping $\mb{x} \mapsto \mb{\phi}(\mb{x})$ is guaranteed by Mercer's theorem, given that the kernel function 
fulfills certain conditions~\cite{Boser:1992,Mercer415}, and in general the feature space $F$ may be an even infinite dimensional 
Hilbert space. For an in-depth exposition on kernel techniques in machine learning see for example~\cite{schoelkopf2002}.
A  common and in many applications successful choice~\cite{Hsu2010} is the Gaussian radial basis function (RBF) kernel:
\begin{equation}\label{kernel}
K(\mb{x}_i, \mb{x}_j) = e^{-\gamma |\mb{x}_i - \mb{x}_j |^2}.
\end{equation}
The width of this kernel function is controlled by the value of $\gamma$ which is the second hyper-parameter that must be set 
to a sensible value \textsl{before} the training of the SVM. We note that the RBF kernel only contains one parameter and that 
the components of the data vectors are added quadratically. Therefore, it is useful to normalize the individual components of the 
input vectors in an appropriate way. Such a scaling ensures that all components of the input vectors may contribute equally.
For this paper we use the range between minimum and maximum value for each component. The training data from Eq.~\ref{for:tra} is replaced by
\begin{eqnarray}
\mb{x}_i=(x_{i,0},\dots,x_{i,n}) \rightarrow \mb{x'}_i=(c_0 x_{i,0},\dots,c_n x_{i,n}),\qquad i=1\dots N,\\
c_k=1/(\max_{i=1\dots N} x_{i,k}-\min_{i=1\dots N} x_{i,k}),\qquad k=1 \dots n,\label{for:scale}
\end{eqnarray}
where $N$ is the size of the training sample and $n$ the dimension of the data vectors~$\mb{x}$. 
Later, in the test phase, the identical scaling constants \ref{for:scale} must be applied to the test data.

\subsection{Probabilistic output}\label{sec:prob}
The SVM described so far is a binary classifier with a \textsl{yes} or \textsl{no} output. In many cases
a posterior probability $P$ that quantifies the belief in the class label is useful and offers an easier interpretation. 
In Section~\ref{sec:tuning}, a probability cut, $P>P_0$, is used to modify the signal-to-background ratio and the total 
number of selected events. Such a classification probability can be estimated~\cite{platt99,wu2009} by fitting a sigmoid model 
to the training data $(y_i,\mb{x}_i)$:
\begin{equation}\label{for:sigmoid}
P(y=1|\hat{f})=\left\{  \begin{array}{ll} \frac{\exp(-t)}{1+\exp(-t)}  &: t \equiv A+B\hat{f} \geqslant0\\ 
                                                          \frac{1}{1+\exp(t)}             &: t < 0
                               \end{array}\right. \;\;,
\end{equation}
where the decision value $\hat{f}$ is given in \ref{for:deci}. 

In general, especially for the non-linear SVM, the result will be biased if the SVM training data itself is used for the fit. 
In LIBSVM the bias is avoided by a five-fold cross-validation\footnote{The procedure of k-fold cross-validation splits the training data randomly into k equal sized subsamples. One subsample is retained for validation while the remaining k-1 subsamples are used for training. The training is repeated k times with changing roles such that each subsample is used exactly once for validation.}. 
It is important to note that a strictly decreasing function of the decision value, as \ref{for:sigmoid}, does not change the order of any sequence of decision values. 
Since the cross-validation increases the computational burden we do not calculate the probabilities during the parameter scan but only for presenting the final results.

% !TEX root = svm_paper.tex

\section{Hyper-parameter tuning}\label{sec:tuning}
The two parameters $C$ and $\gamma$, introduced in the previous section, must be set to sensible values before the training of the support vector machine.
These values are crucial for the performance of the algorithm, and different strategies to optimize the parameter choice are possible.
The easiest approach is a simple grid search. A two dimensional grid is defined, at each grid point the SVM training is performed on a training dataset, and the trained machine is applied to an independent test data sample where some performance measure is evaluated. Eventually, the $(C,\gamma)$-pair with the highest performance index is used. We first consider what could be an appropriate performance measure for a new physics search and describe then the tuning algorithm in some detail.
 
\subsection{Performance measures}

\paragraph{Machine learning performance measures.}
From a machine learning perspective, a natural performance measure describes how well a classifier separates the two distinct classes. 
On a sample of test data we know the true class labels. There are 2x2 categories formed by the true label $y\in \{-1,1\}$ and the label $\hat{y}\in\{-1,1\}$ estimated by the SVM. The relative amount of test data in these categories can be used to quantify the performance of a machine learning algorithm.
Typical ML measures are the \textsl{accuracy} which gives the percentage of correctly predicted labels, or the {\textsl{precision} which, in our case, is the percentage of correctly predicted signal events. 
%Other quantities are the \textsl{recall}, the filter efficiency for signal events, or the \textsl{F score} calculated as the harmonic mean of \textsl{precision} and \textsl{recall}.
Another frequently used measure is the \textsl{AUC}, the area under the receiver operator curve (ROC).
The ROC curve shows the background rejection (false positive) against the signal efficiency (true positive) at various threshold values
of the decision function.
%Accuracy  = (a+d)/(a+b+c+d) = correct predicted/all - equal cost for all errors
%Precision = a/(a+c)         = \%true in predicted signal
%Recall    = a/(a+b)         = \%found from input signal - filter efficiency
%FMeasure  = 2*P*R/(P+R)     = F score - harmonic mean of precision and recall
%ROC and AUC=area under ROC calculted on order statistic - independent on f -> p(f) if p monotonously increasing
% decision function; sort_f (ty,f) 
% for ty in sort_f (ty,f):
%    if ty>0:tp++
%    elseif ty<0:
%         fp++
%         roc+=tp
%auc=roc/tp/fp
While the use of these and other performance measures is common also in HEP~\cite{Hocker:2007ht}, we will follow a different 
approach to optimize the SVM.

\paragraph{Physics motivated performance measures.} The maximum number of correctly classified events is of secondary importance for a HEP search. 
There is a much more physically and statistically motivated measure: the discovery significance. Optimizing the significance is a common procedure in HEP.
Typically the search area is optimized for a statistically relevant signal to background ratio that allows to prove or reject a certain  hypothesis. 
Here, we consider the case of a cut-and-count analysis for which several significance estimators are commonly used \cite{CCGV2011}\cite{Cousins2008}. Optimizing a certain, statistically motivated, figure of merit is common practice in HEP to select different ML algorithms or different sets of input variables.
The new insight of this paper is that such a procedure can successfully be applied in the stage of model selection, \ie during the hyper-parameter tuning. 
%The different pairs of $(C,\gamma)$ values represent different experimental setups and we must consider  the question how to find the optimal experimental design. 
\paragraph{\textsl{Asimov} estimate.} 
The exact numerical  calculation of the statistical significance may become computationally costly.
A well performing estimate for the discovery significance has been given in  \cite{CCGV2011}.
For the case of Poisson distributed background and signal events ($s$,$b$) with background uncertainty $\sigma_b$ 
the approximated median discovery significance becomes
\begin{equation} 
\label{asimovz}
Z_{A}=\left[2\left((s+b)\ln\left[\frac{(s+b)(b+\bsvar)}{b^2+(s+b)\bsvar}\right]-\frac{b^2}{\bsvar}\ln\left[1+\frac{\bsvar s}{b(b+\bsvar)}\right]\right)\right]^{1/2}.
\end{equation}
\subsection{Hyper-parameter search}\label{sec:hyper}
The SVM with RBF kernel requires two hyper-parameters:  $C$ (Sec.~\ref{sec:over}) and $\gamma$ 
(Sec.~\ref{sec:nonlin}). In addition to the hyper-parameters, the number of selected signal and 
selected background events $(s,b)$ depends on the probability cutoff $P_0$ (Sec.~\ref{sec:prob}), 
or a corresponding decision value $\hat{f_0}$. The easiest algorithm to find the optimal values
for these parameters is a grid search.  At each point of a logarithmically spaced grid in $(C,\gamma)$ 
a SVM is trained and, on an independent test dataset, the Asimov significance $Z_A$ is calculated 
as function of $P_0$. In general, the best cut $P_0$ is selected as the value with the highest significance. 
To avoid artificially high significance values due to statistical fluctuation of a small signal at very 
low values of $b$, a further requirement of at least 5 signal events is applied. 
While conceptually the plain grid search is sufficient to find good values for $(C,\gamma)$, computationally 
it may be advantageous to use a more refined algorithm for the hyper-parameter tuning. The details of the 
iterative algorithm used for the results in this paper are given in Appendix~\ref{app:itergrid}.
 
% !TEX root = svm_paper.tex

\section{Case studies}\label{sec:cases}
%\subsection{BDT}
\subsection{Performance comparison on a toy model} 

Comparing speed and classification performance of different classifiers is not always straightforward.
In order to have simple and well defined conditions, we start with a toy model and 
compare our SVM-HINT framework with a BDT and an SVM, both implemented with the TMVA library. 
The toy model includes the following variables $V_i$ generated with the random numbers $x_i$ (the \textsl{tilde} means sampled from):
\begin{equation}
\begin{array}{rcl}
V_1 = &\sin(x_1);\qquad & x_1 \sim g(x_1|a,b)\\
V_2 = &x_2;\qquad &x_2 \sim \exp(-x_2/c)\\
V_3 = &x_3;\qquad &x_3 \sim g(x_3|d,e)\\
V_4 = &\sqrt{x_4};\qquad &x_4 \sim \exp (-x_4/f)
\end{array}
\end{equation}
where $g(x|\mu,\sigma)$ is a Gaussian distribution with mean $\mu$ and width $\sigma$, %= \frac{1}{\sqrt{2}\sigma}\exp\left(-\frac{(x-\mu)^2}{2\sigma^2}\right)$
and $a,b,c,d,e,f>0$ are constants with different values for signal and background samples. 
%The toy model creates weighted events such that signal events get on average a 50 times smaller weight which is similar to the detailed example in the following section. 
This model does not have any hidden correlation between the variables and each ML algorithm needs only to find a 
set of independent optimal cuts. Due to its simplicity, the toy model enables us to generate large quantities of 
events to study the training time as function of the training sample size for the different codes. 

As explained in Sec.~\ref{sec:tuning}, SVM-HINT provides a hyper-parameter search. The hyper-parameter 
search is performed beforehand and is not part of the timing performance study.
The SVM implementation provided by the TMVA library lacks such an automated hyper-paramter search. We therefore apply %use the RBF kernel for this case with
the same hyper-parameter values as obtained by the SVM-HINT tuning algorithm. 
The out-of-the-box performance of the TMVA-BDT cannot compete in most cases with the automatically tuned SVM-HINT. 
There is a trade-off between classification performance and time consumption of the BDT
which can be controlled by an appropriate choice of the BDT parameters, e.g.\ the number of trees, minimum node size, and 
cut values, as introduced in Sec.~\ref{BDT}.   
For comparing the training times, we follow the strategy to optimize the BDT parameters manually\footnote{Configuration files are available at \cite{ourgithub}.}
to accomplish a similar discovery significance as achieved with the  SVM-HINT. This allows us to compare the time consumption of equally performing algorithms. Blindly optimizing for maximal performance on a fixed evaluation training sample could otherwise produce slow, over-sized trees and would place a disadvantage on the BDT implementation.

\begin{figure}[!h] 
\centering
\includegraphics[width=0.6\textwidth]{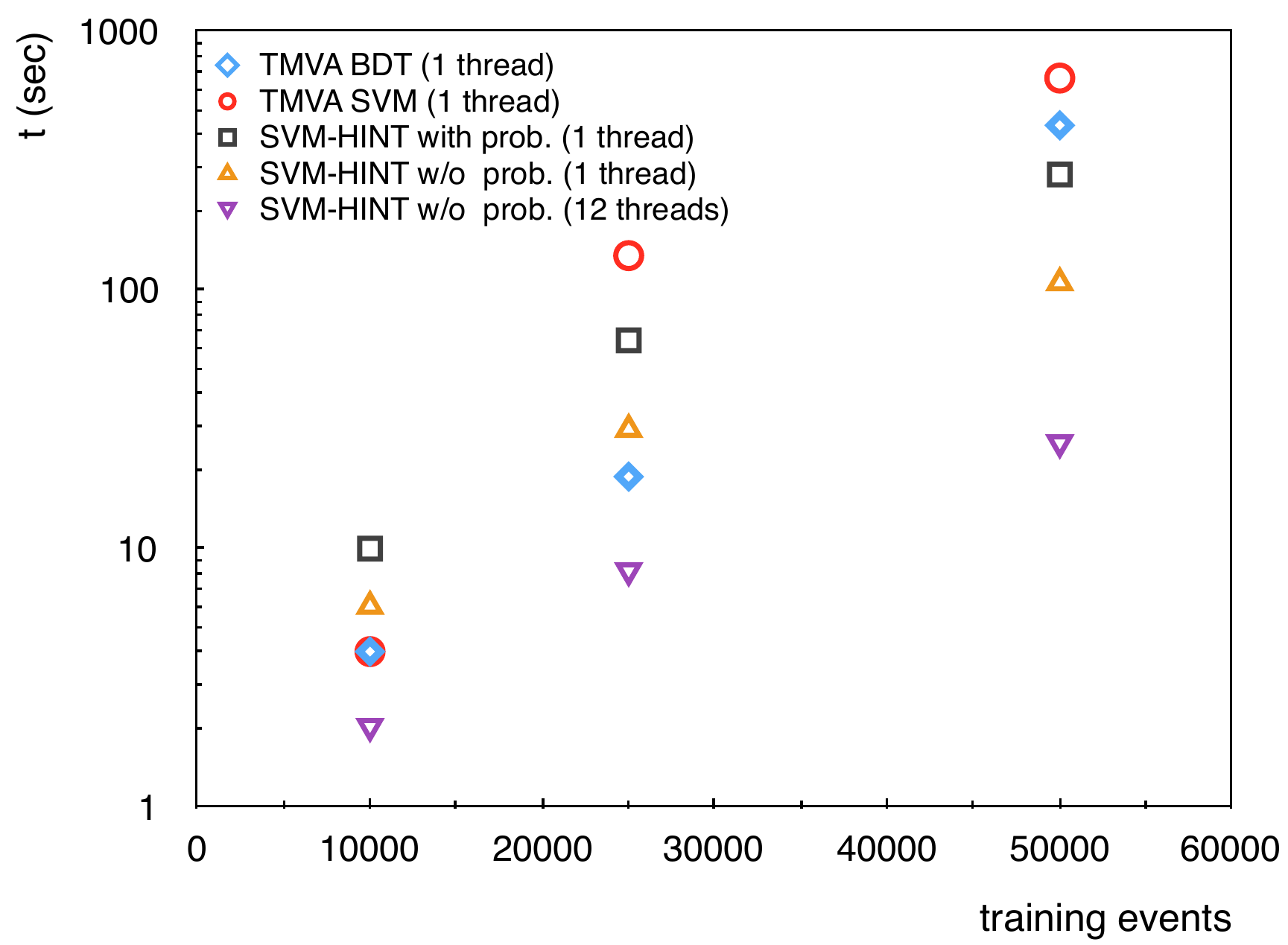}
\caption{The timing performance of the classifiers are compared on a computer with two Intel\textsuperscript{\textregistered} Xeon\textsuperscript{\textregistered} E5-2440 CPUs and 12 physical threads running at 2.40 GHz clock speed. Number of threads used by each classifier implementations are stated within parentheses in the legend.
}
\label{fig:timecomp}
\end{figure}

Fig.~\ref{fig:timecomp} shows the results for the different classifiers. At low numbers of training events the BDT performs 
better than the other single-threaded classifiers. With increasing training sample size the number of trees needed to achieve 
a competitive classification performance becomes larger with negative impact on the training time.
%The higher the number of training events, the higher the number of trees becomes we need to achieve a competitive classification performance 
%At higher numbers of training events we have to increased the number of trees to achieve a competitive classification performance which resulted with slower training times for the BDT. 
The TMVA-SVM does not scale well in terms of timing performance and it performs poorly on bigger samples. Overall the SVM-HINT performs similar or better compared to TMVA-BDT and TMVA-SVM. In addition, the SVM-HINT can efficiently take advantage of multi-core architectures. Naturally, the multi-threaded performance of the SVM-HINT with 12 threads outperforms the other implementations.

% !TEX root = svm_paper.tex

\subsection{Third generation supersymmetric partner search}\label{sec:stop}
\paragraph{Monte Carlo samples}

As a real-world physics example we consider a search for the supersymmetric partner of the top quark, called top squark, at the LHC.
The search is designed for the case of direct top-squark pair production with subsequent decay of the top squarks 
into the lightest supersymmetric particle (LSP) and a top quark. Several searches for such a scenario have been performed at 
TeVatron as well as at the LHC~\cite{Abazov2004147,Abazov:2012cz,Abazov:2007ak,Aaltonen:2010uf,Acosta:2003ys,Aad:2015pfx,Chatrchyan:2013xna,Berggren:2015qua}. 
After preselection of the data, the remaining dominant background is given by top-antitop (\ttbar) quark production. Top quarks 
decay to almost 100\% to a a b quark and a W boson, with the latter decaying either to two quarks or to a lepton and a neutrino.
When requiring one lepton in the final state, we mainly expect to select semi-leptonic decays of top quarks (where one W boson decays
to two jets and the other to a lepton and a neutrino), but dileptonic top decays (where both W bosons decay to a lepton 
and a neutrino) might be selected as well, if one lepton is not identified (lost) for various reasons.
A leading order simulation is sufficient for our purpose and we only take into account NLO results for the total cross sections of 
signal~\cite{Borschensky:2014cia} and background~\cite{Campbell:2010ff,Nadolsky:2008zw} processes. 
{\sc Pythia6}~\cite{Sjostrand:2006za} is used for the event simulation and  {\sc Delphes3}~\cite{Ovyn:2009tx} to model 
the detector response. The detector model is a \textsl{combined} ATLAS and CMS detector, as it had been used for the 2013 Snowmass effort~\cite{Anderson:2013kxz}. 

We categorize the physical variables with respect to their mathematical complexity as high-level and low-level variables. 
The low-level variables consist of basic properties of the reconstructed physics objects measured by the detector, while 
the high-level variables are constructed from the low-level variables using physical insight to improve the classification 
performance. The physics objects are jets and leptons, where lepton is used as generic term for electrons and muons. 
For simplicity, we do not consider tau leptons since their experimental reconstruction is more complex.

Low-level variables are the transverse momentum \pt and the pseudorapidity $\eta$ of the single lepton, of the four 
highest-\pt (called `leading') jets, and of the leading b-quark jet. 
In hadron collider experiments the missing energy perpendicular to the beam direction, \ETslash, is commonly reconstructed 
as an independent variable. It is therefore treated as a low-level quantity, as well as \HT, the scalar sum over the 
transverse momenta of all preselected jets. In many SUSY models, we expect large \ETslash due to the LSP, which
is expected to be neutral and weakly interacting and will therefore not be detected. As SUSY particles are heavy, we also expect
a large amount of energy in the detector leading to large \HT. In addition, the multiplicities of jets  ($n_{\mathrm{jet}}$) and b-quark 
jets ($n_{\botq \mathrm{jet}}$) are included.

As high-level variables we consider the following variables: the transverse mass \MT, defined as 
$\MT = \sqrt{ 2\,  p_{\mathrm{T},\lep}\,  \ETslash\,  (1 - \cos \Dp (\lep,\ETslash))}$, can be used to 
suppress the background from W-boson production, as \MT of leptonic W decay events does not exceed the 
the W mass. Dileptonic \ttbar-events with one lost lepton are an important background since the lost 
lepton mimics large missing energy from the LSP. The \MTtW variable~\cite{MT2W} is constructed 
exploiting the knowledge of the \ttbar-decay kinematics to separate such events. Top-squark production is a high-mass process 
with large missing energy. High-mass production is typically related to more centrally distributed particles in the detector, such 
that the \textsl{Centrality}, defined as $\sum_{\mathrm{jets, \lep}} \pt \,/\! \sum_{\mathrm{jets, \lep}} p$, 
can be used to enhance such events. Commonly used relations between the hadronic activity and 
the missing energy are $Y=\ETslash \,/\! \sqrt{\HT}$, often referred to as \ETslash significance, and
the $\HT$-ratio, the normalized hadronic activity in the hemisphere of $\ETslash$.
The last group of variables exploit topological relations between the event particles: 
\dphilw is the angle between the \W boson and lepton,
\drlb is the radial distance between closest lepton and \botq-jet and 
\mlb is the invariant mass of the \botq-jet and the closest lepton.

A compilation of all low-level and high-level variables is given in Table~\ref{tab:varsets}, together with
the definition of four sets of variables which  are considered to investigate the influence of the variable 
multiplicity and complexity in the multivariate analysis. We define one set containing all variables, one
using only low- or high-level variables, respectively, and a subset of two low-level and two high-level variables
with relatively large separation power. 

\begin{table}[!h]
\caption{\label{tab:varsets} 
Summary of all low-level and high-level variables used in the analysis. 
Set 1 includes all variables. Set 2 and set 3 consist of low- and high-level variables, respectively. 
Set 4 is a smaller subset of high- and low-level variables. }
\vspace{2em}
\centering
\begin{tabular}{|c|l|c|c|c|c|}
\hline
&    Variable                  & Set 1       & Set 2       & Set 3       & Set 4\\
\hline
\parbox[t]{2mm}{\multirow{10}{*}{\rotatebox[origin=c]{90}{low-level}}}  
&$p_{\mathrm{T},\lep}$         & \textbullet & \textbullet &             & \\
&$\eta_{\lep}$                 & \textbullet & \textbullet &             & \\
&$p_{\mathrm{T},jet(1,2,3,4)}$ & \textbullet & \textbullet &             & \\
&$\eta_{jet(1,2,3,4)}$         & \textbullet & \textbullet &             & \\
&$p_{\mathrm{T},\botq\, jet1}$ & \textbullet & \textbullet &             & \\
&$\eta_{\botq\,jet1}$          & \textbullet & \textbullet &             & \\
&$n_{jet}$                     & \textbullet & \textbullet &             & \\
&$n_{\botq\,jet}$              & \textbullet & \textbullet &             & \\
&$\ETslash$                    & \textbullet & \textbullet &             & \textbullet \\
&$\HT$                         & \textbullet & \textbullet &             & \textbullet \\
\cline{1-2}%--------------------
\parbox[t]{2mm}{\multirow{10}{*}{\rotatebox[origin=c]{90}{high-level}}}  
&$\MT$                         & \textbullet &             & \textbullet & \textbullet \\
&$ \MTtW$                      & \textbullet &             & \textbullet & \textbullet \\
&$\dphilw$                     & \textbullet &             & \textbullet & \\
&$\mlb$                        & \textbullet &             & \textbullet & \\
&Centrality                    & \textbullet &             & \textbullet & \\ 
&$Y$                           & \textbullet &             & \textbullet & \\
&$\HT$-ratio                   & \textbullet &             & \textbullet & \\
&$\drlb$                       & \textbullet &             & \textbullet & \\
&$\dphimin$                    & \textbullet &             & \textbullet & \\ 
\hline
\end{tabular}
\end{table}

\paragraph{Analysis strategy}

In order to reduce the time required for training and optimization, a baseline selection, summarized in Table~\ref{tab:bcut}, 
is applied to the signal and background samples. Figure~\ref{fig:varplots} shows the distribution of signal and background
for two low-level and two high-level variables, \HT, \ETslash, \MT and \MTtW after the baseline selection, normalized to the expected 
luminosity at the end of the LHC run in the year 2023, corresponding to 300\fbinv. The background is several orders of magnitude
higher than the signal, and the distributions of signal and background are quite similar due to their similar kinematics.  

\begin{figure}[!h]
\centering
\subfigure{\includegraphics[width=0.475\textwidth]{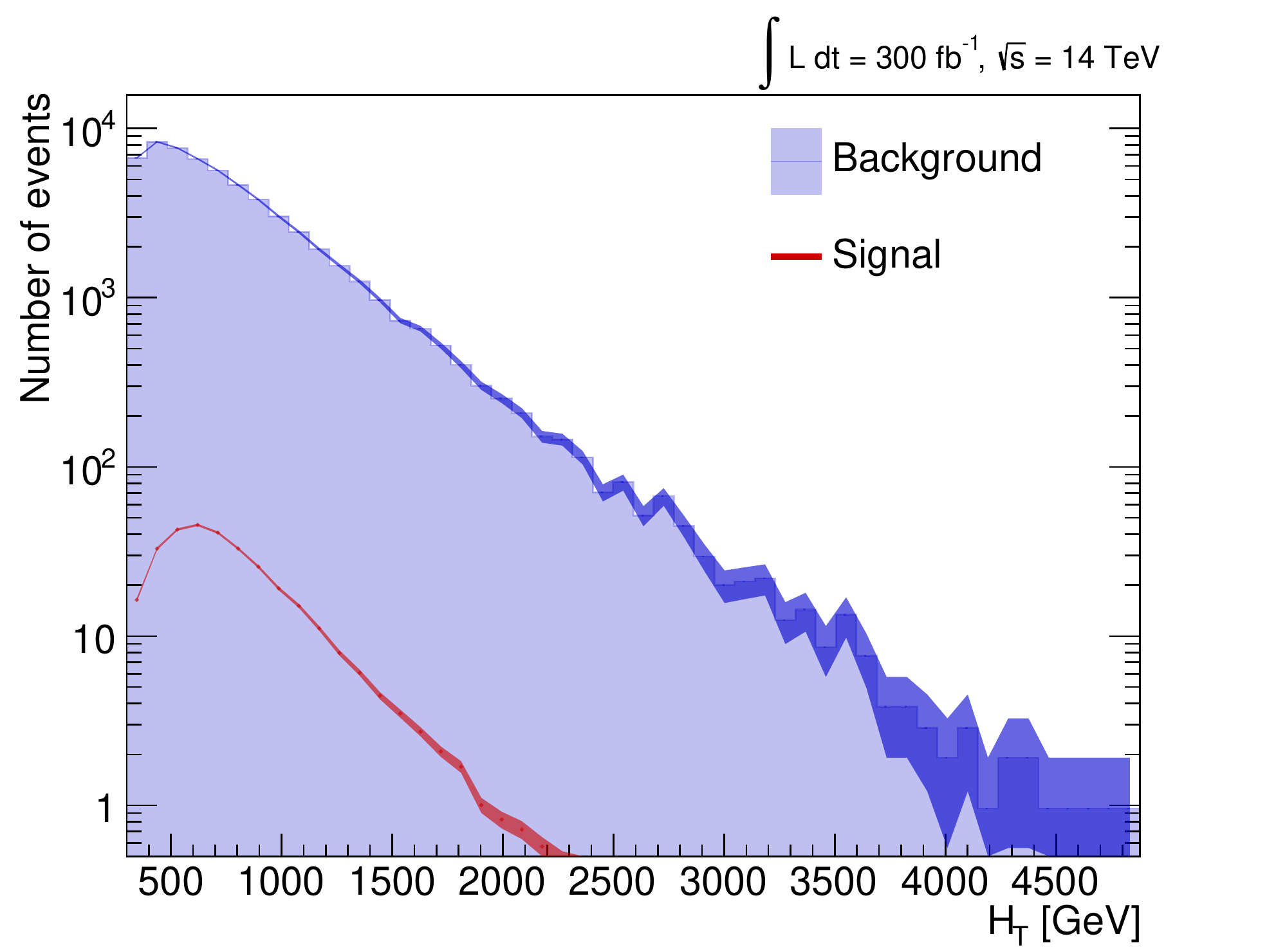}   \label{fig:ht}}
 \subfigure{\includegraphics[width=0.475\textwidth]{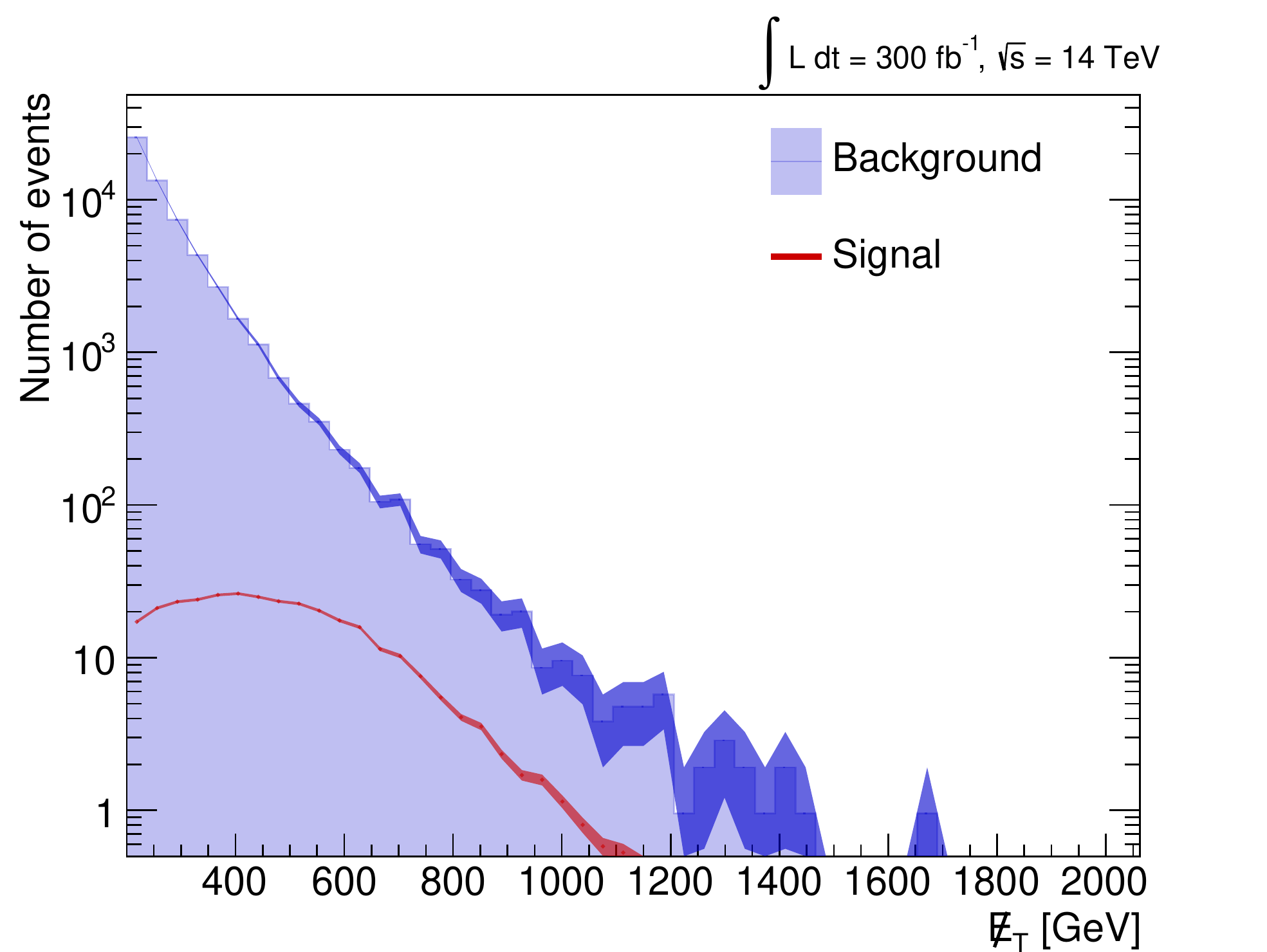}  \label{fig:met}}
\subfigure{\includegraphics[width=0.475\textwidth]{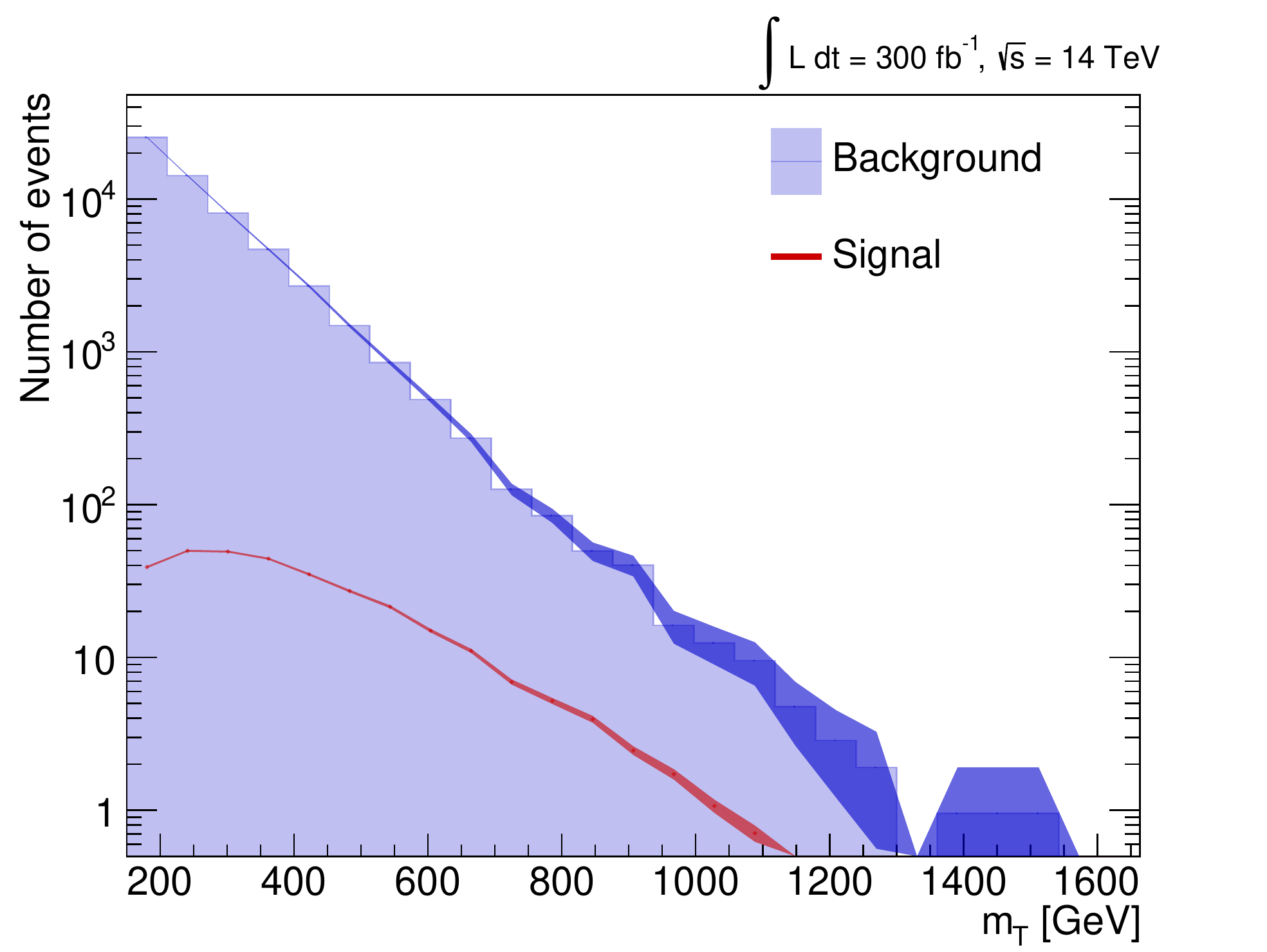}     \label{fig:mt}}
   \subfigure{\includegraphics[width=0.475\textwidth]{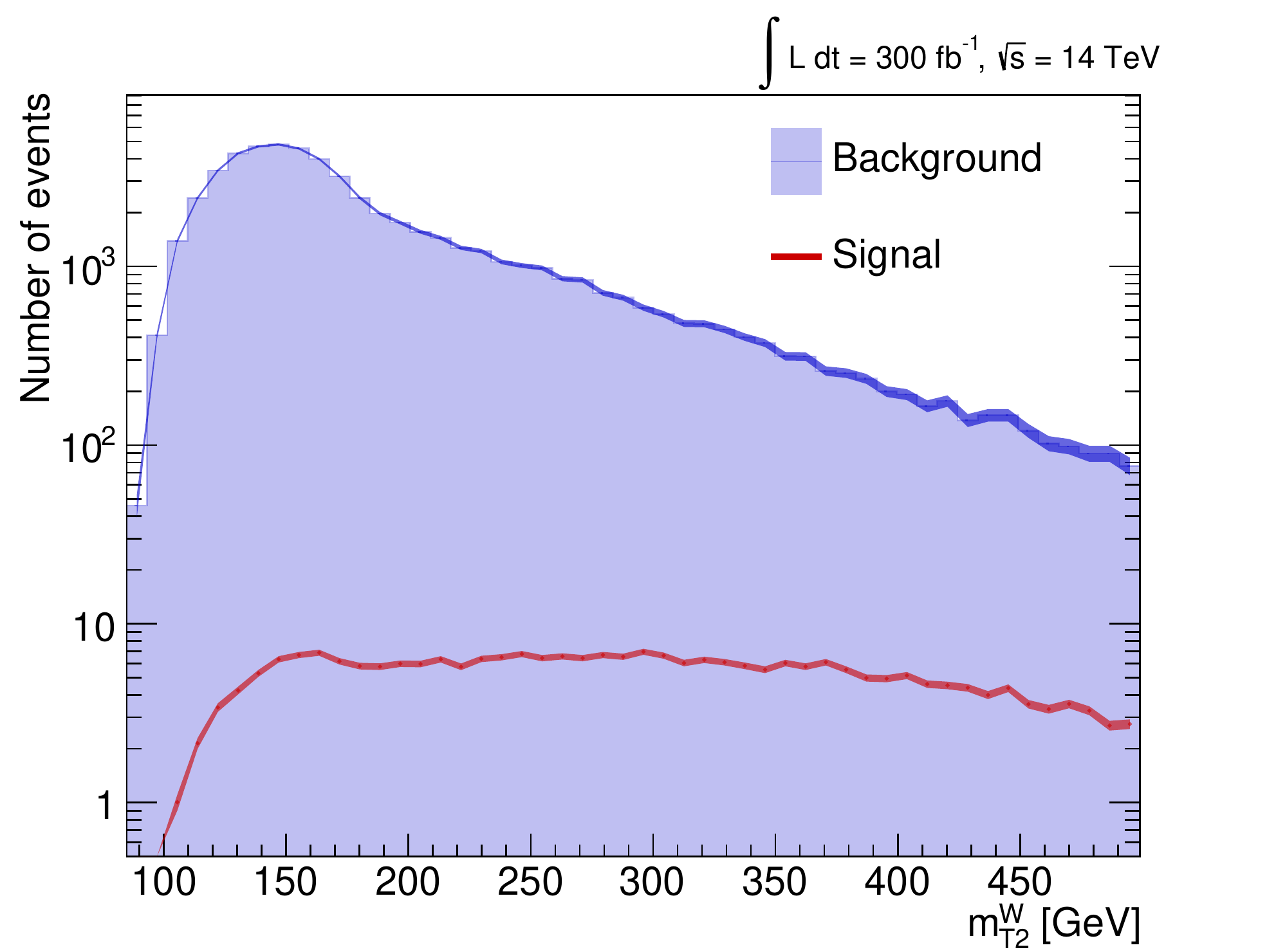}  \label{fig:mt2w}}
\caption{The distribution of signal (red line) and background (blue filled histogram) after the baseline selection for two low-level and two high-level variables that are used in the analysis: \HT, \ETslash, \MT and \MTtW. The y-axis shows the number of events 
(normalized to the integrated luminosity), and the x-axis shows the variable value for a given bin. Statistical errors 
are represented by transparent bands.}
\label{fig:varplots}
\end{figure}

The samples are separated into three independent subsamples: training, test and 
evaluation sample. Each classification method is optimized over the training and test samples and the best-performing configuration 
is applied to the independent evaluation sample for the final performance assessment.

\begin{table}[!h]
\caption{\label{tab:bcut} Top-squark search: List of baseline selection requirements}
\vspace{2em}
\centering
\begin{tabular}{|rcl|}
\hline
$|\eta_{\lep,\,\mathrm{jet}}|$ & $ < $ & $ 2.4$ \\
$p_{\mathrm{T},\lep}$ & $ > $ & $ 30 \GeV $ \\
$p_{\mathrm{T},\mathrm{jet}}$ & $ > $ & $ 40 \GeV $ \\
$p_{\mathrm{T},\mathrm{jet 1}}$ & $ > $ & $ 80 \GeV $ \\
$p_{\mathrm{T},\mathrm{jet 2}}$ & $ > $ & $ 60 \GeV $ \\
$\ETslash$ & $ > $ & $ 200 \GeV$ \\
$\HT$ & $ > $ & $ 300 \GeV$ \\
$n_{jet}$ & $ > $ & $ 3 $ \\
$n_{\botq\,jet}$ & $ > $ & $ 0 $ \\
\hline
\end{tabular}
\end{table}

The TMVA-BDT has been manually trained and tested over 8 different settings for each of the four variable sets in order 
to obtain optimal parameters as described in Appendix~\ref{BDT}, 
while the SVM-HINT is auto-tuned by the iterative grid search, as described in Sec.~\ref{sec:tuning} and Appendix~\ref{app:itergrid}. 
Without modifying the default SVM-HINT settings, the two step grid search hyper-parameter optimization function provides the optimal 
parameters using test and training samples. We calculate the final significance with an independent evaluation sample.

\begin{figure}[!h]
\centering
\subfigure{\includegraphics[width=0.475\textwidth]{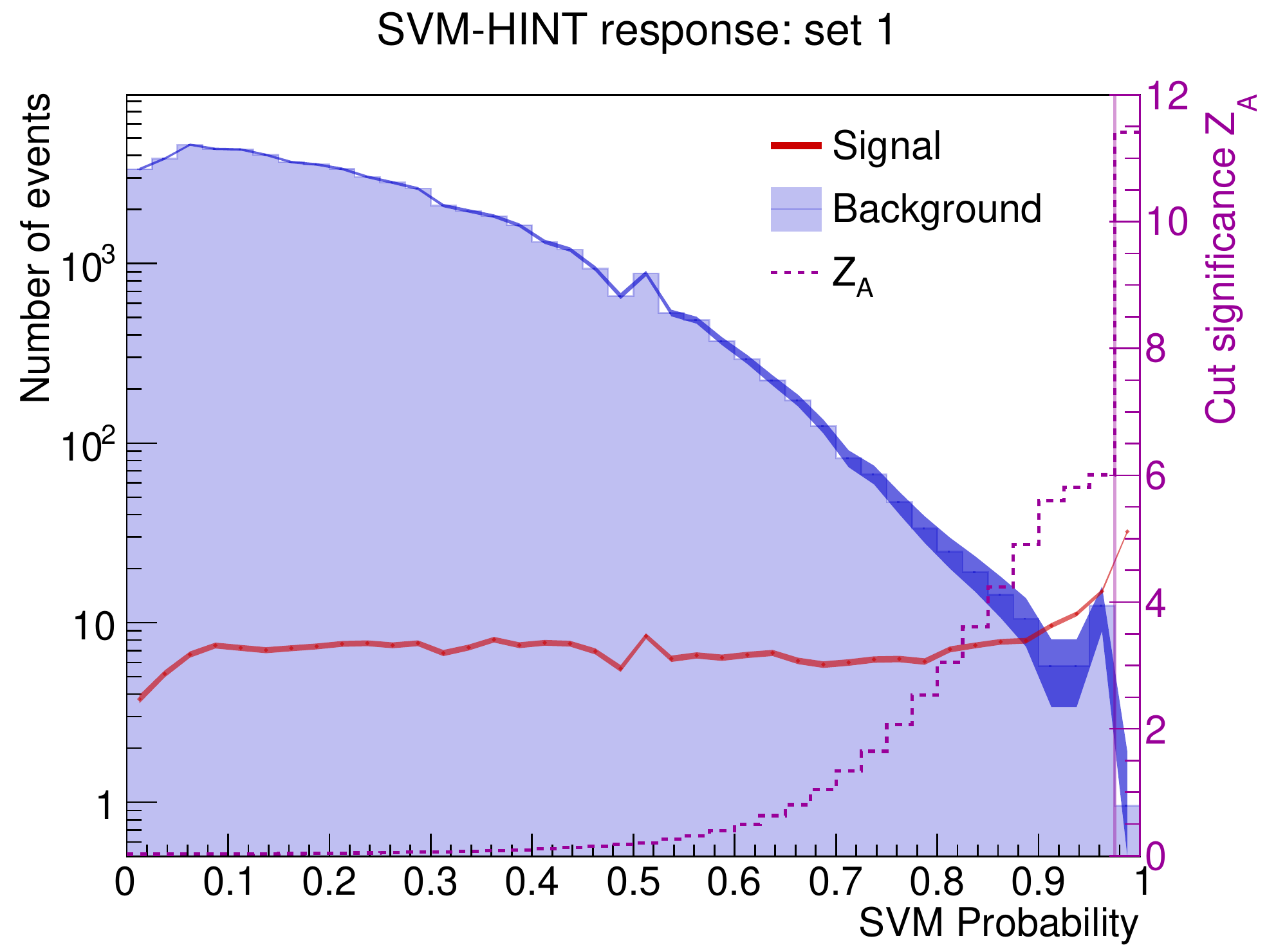} \label{fig:allsvmhint}}
 \subfigure{\includegraphics[width=0.475\textwidth]{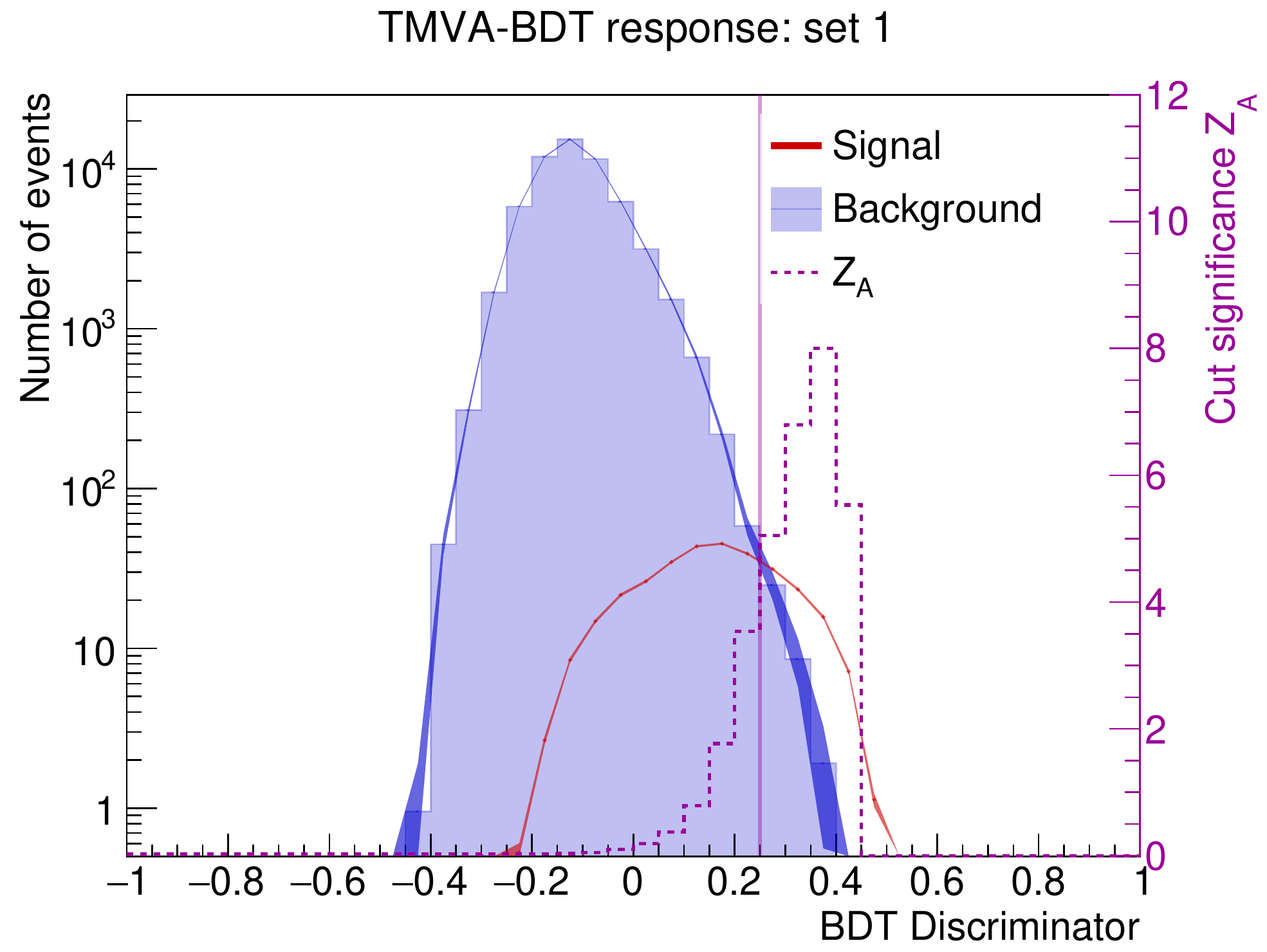} \label{fig:alltmvabdt}}
\subfigure{\includegraphics[width=0.475\textwidth]{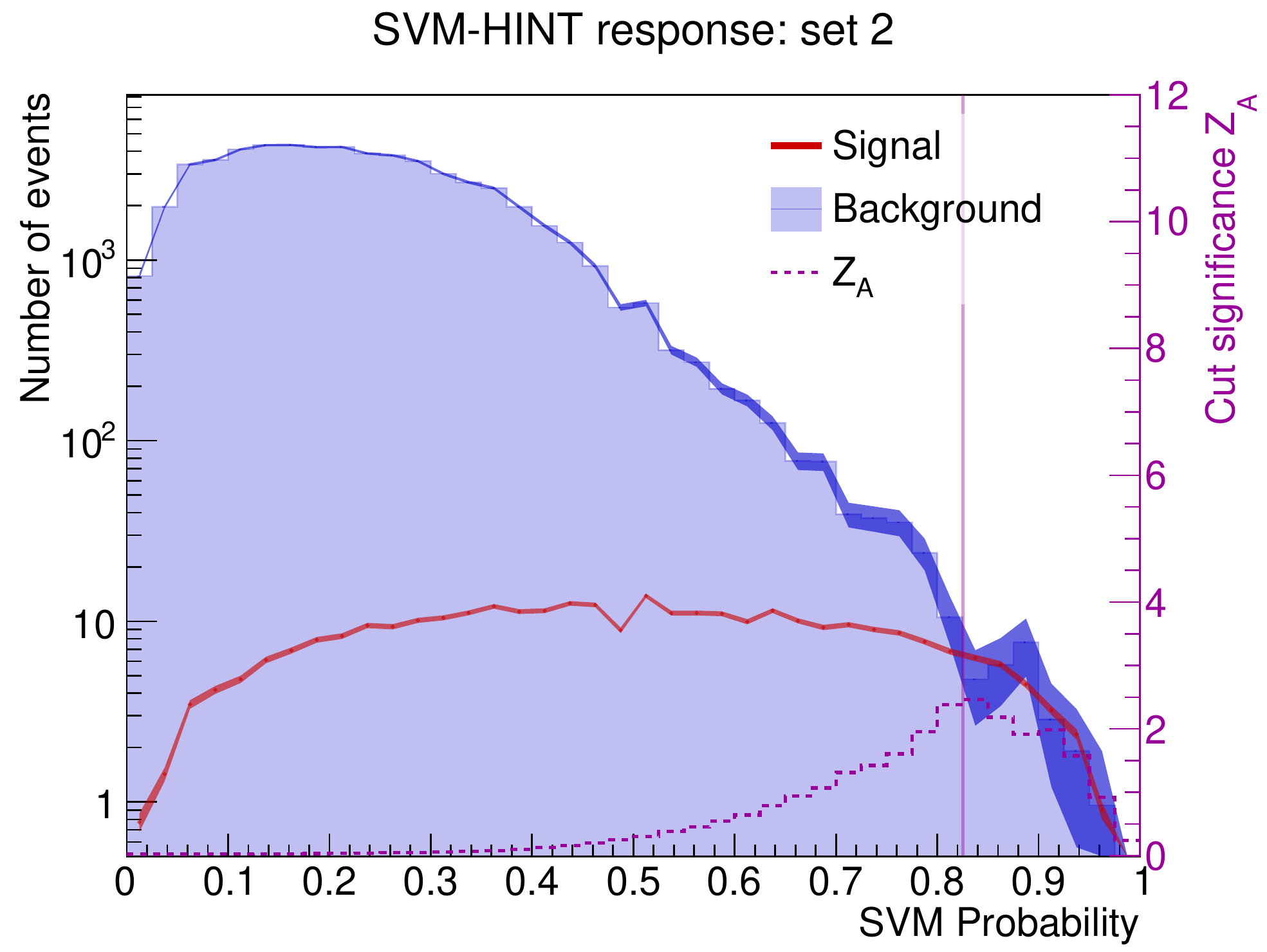}  \label{fig:lowsvmhint}}
 \subfigure{\includegraphics[width=0.475\textwidth]{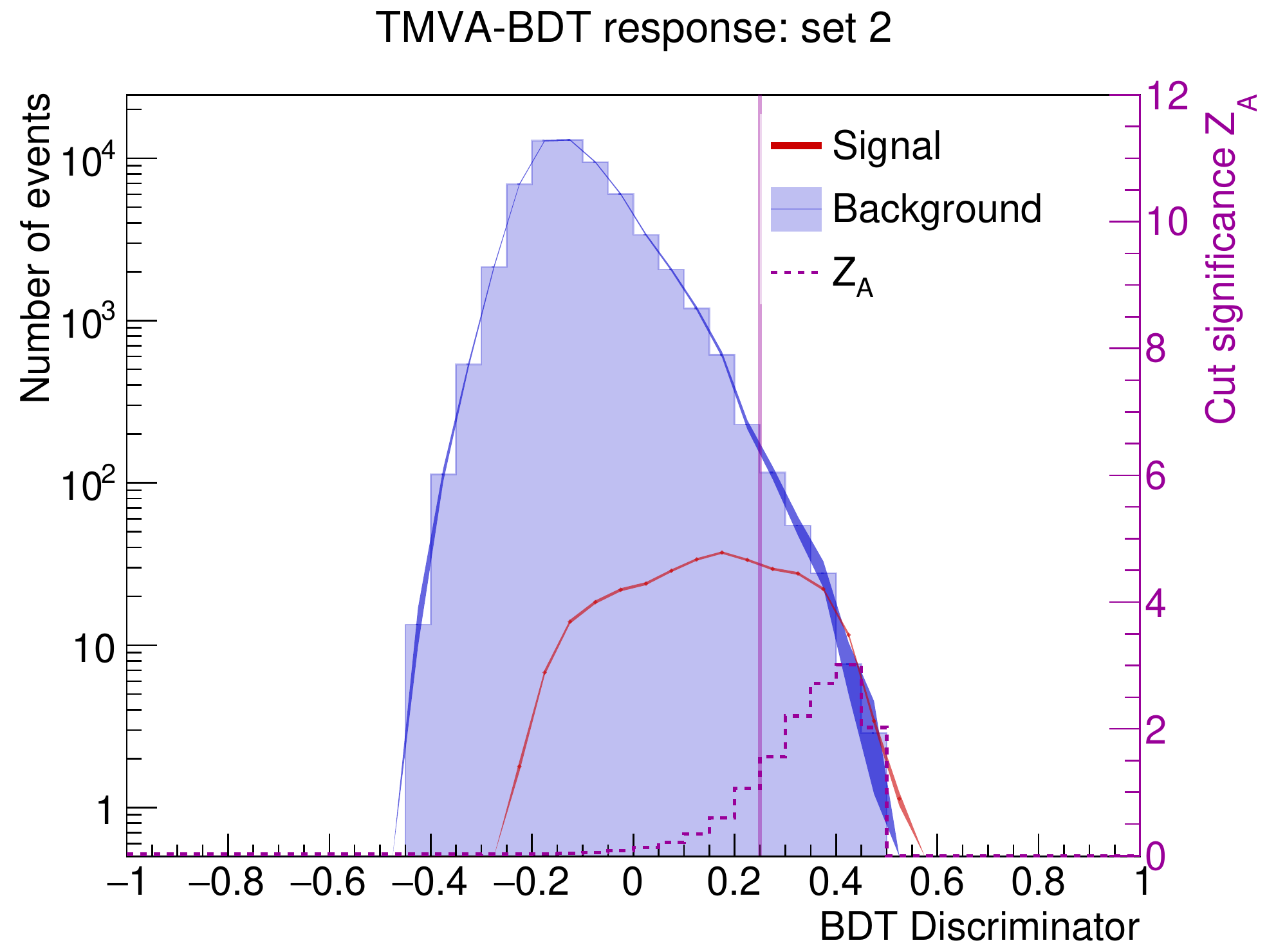} \label{fig:lowtmvabdt}}
 \caption{The SVM-HINT and TMVA BDT responses trained with the variable sets set 1 and set 2, as defined in \ref{tab:varsets}. 
Even though the optimal $Z_{A}$ efficiency information is not available in data, it is included to demonstrate the 
reliability of the optimal discriminator cut output from the classifier implementations. The y-axis on the left shows 
the number of events (normalized to the aimed integrated luminosity), whereas the y-axis on the right shows the Asimov 
significance for the discriminator cut per bin. }
\label{fig:hintsvmtmvares1}
\end{figure}

\begin{figure}[!h]
\centering
\subfigure{\includegraphics[width=0.475\textwidth]{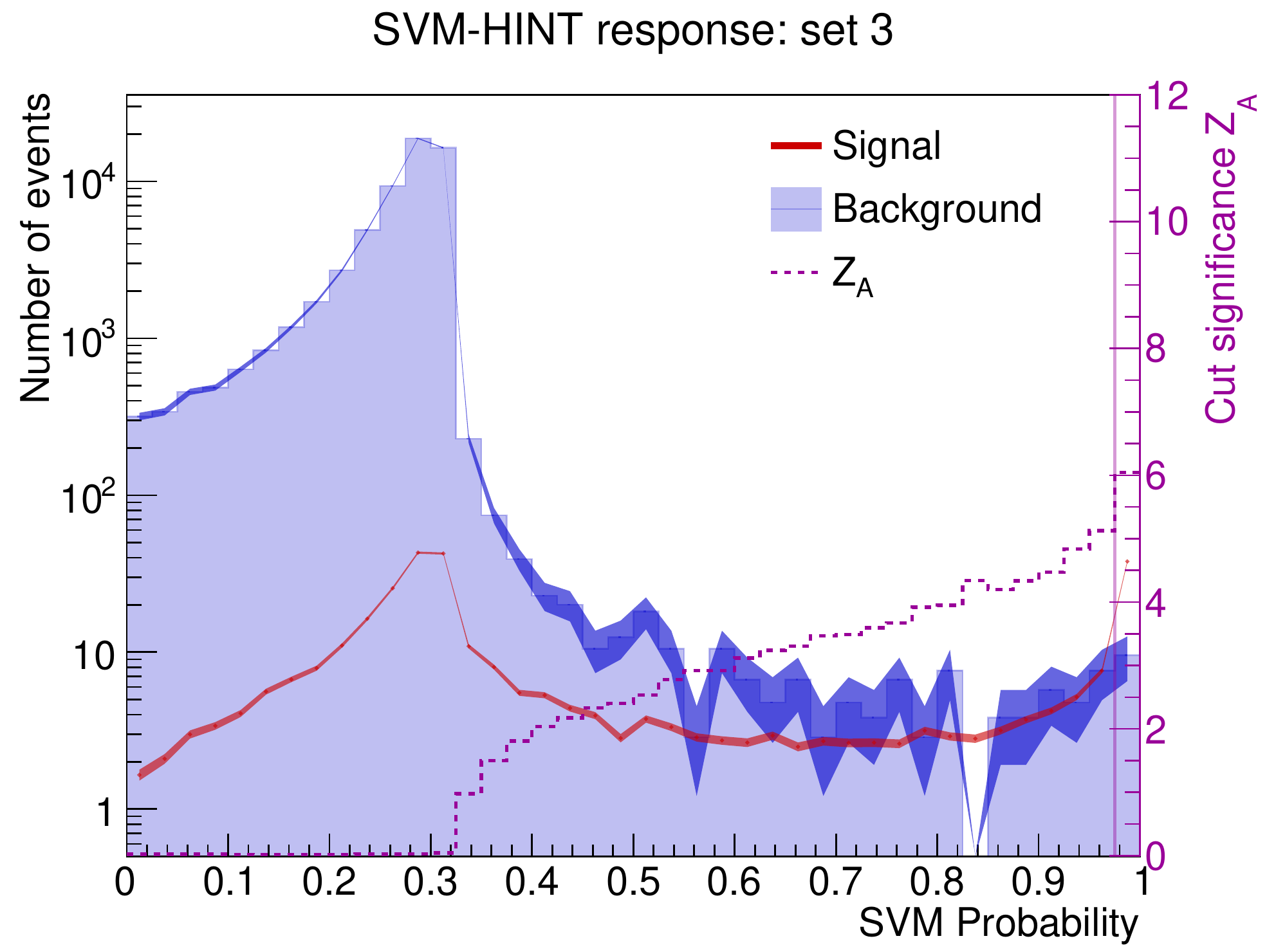} \label{fig:highsvmhint}}
 \subfigure{\includegraphics[width=0.475\textwidth]{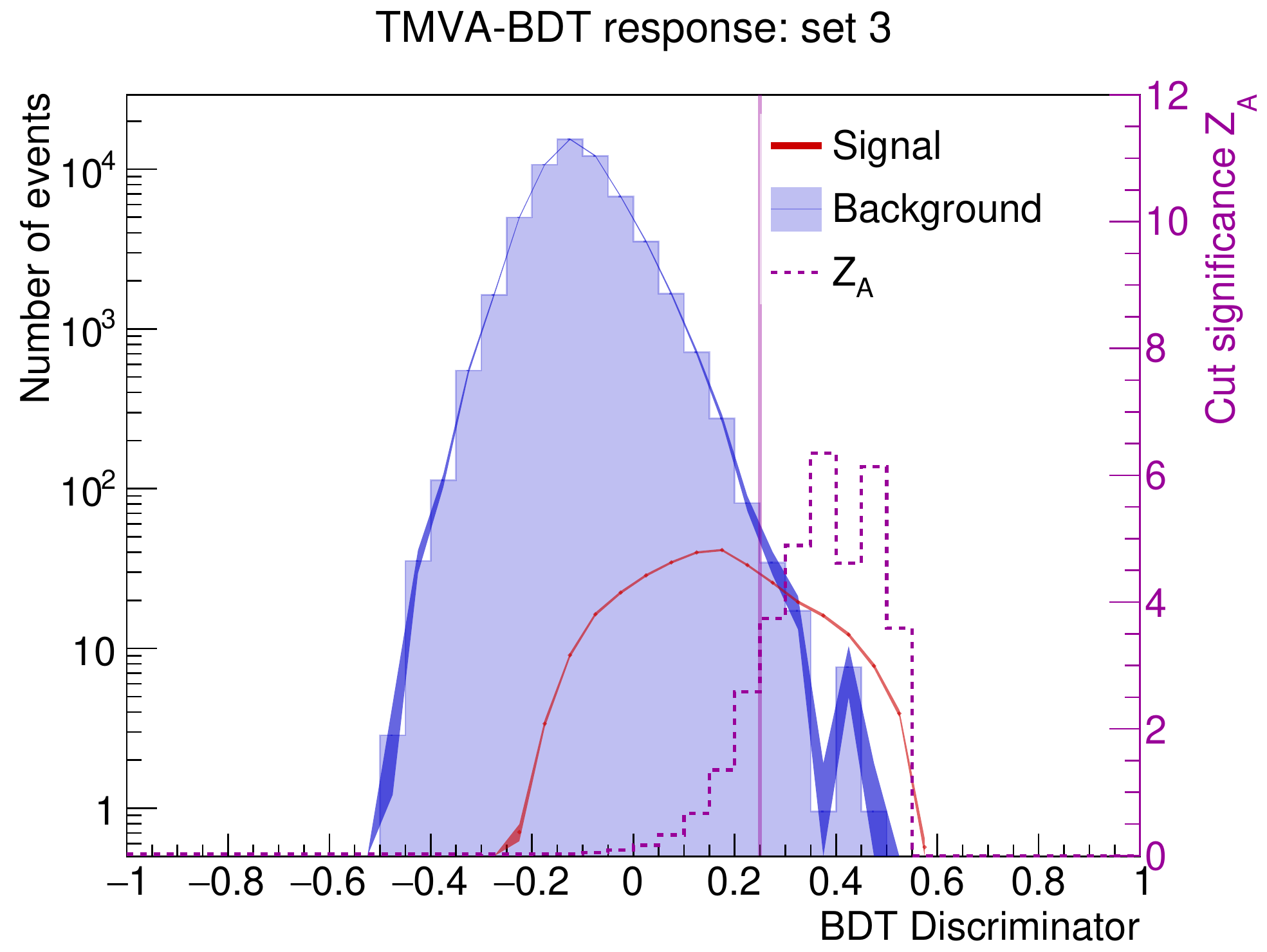}  \label{fig:hightmvabdt}}
\subfigure{\includegraphics[width=0.475\textwidth]{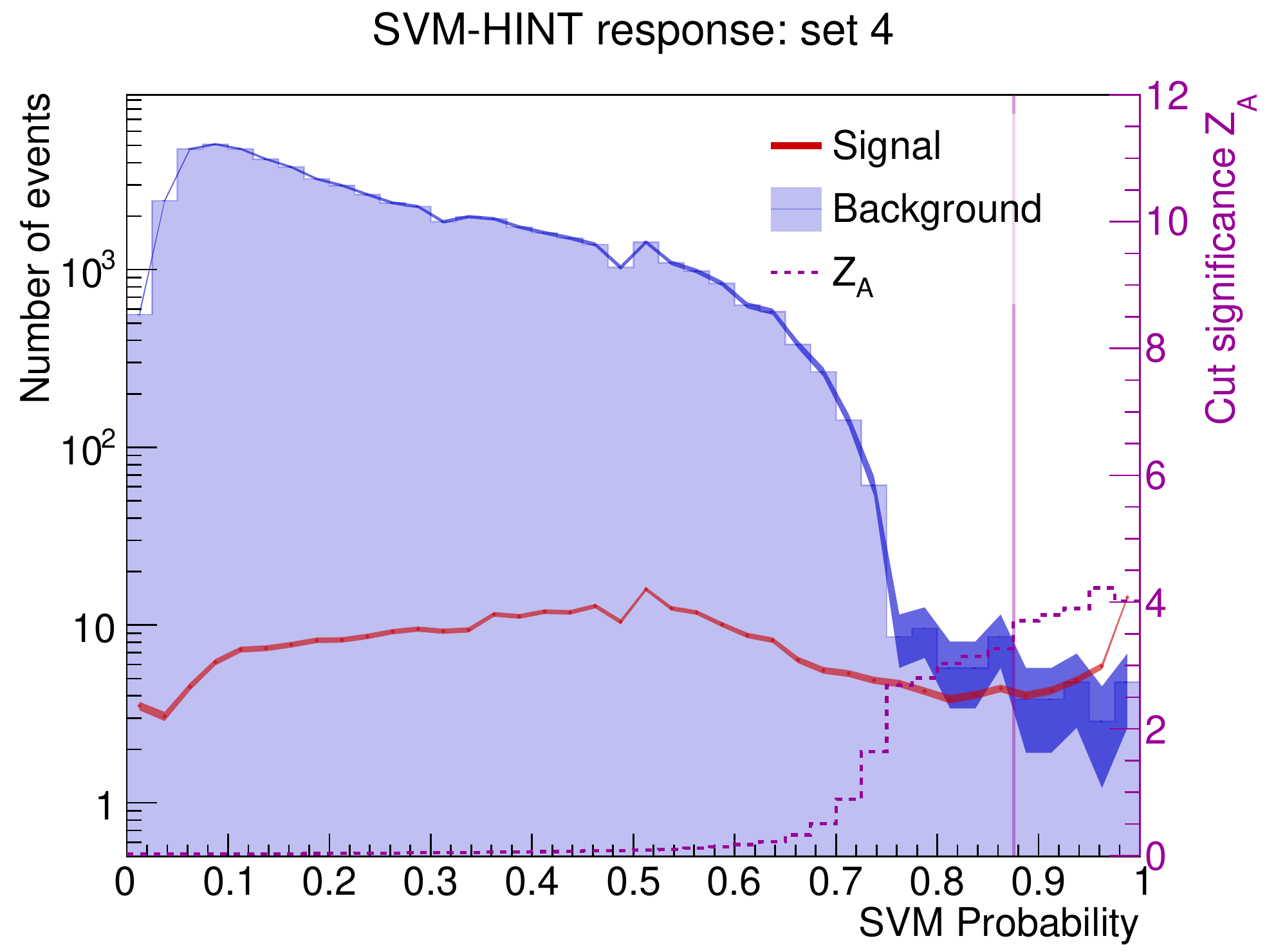}  \label{fig:subsvmhint}}
   \subfigure{\includegraphics[width=0.475\textwidth]{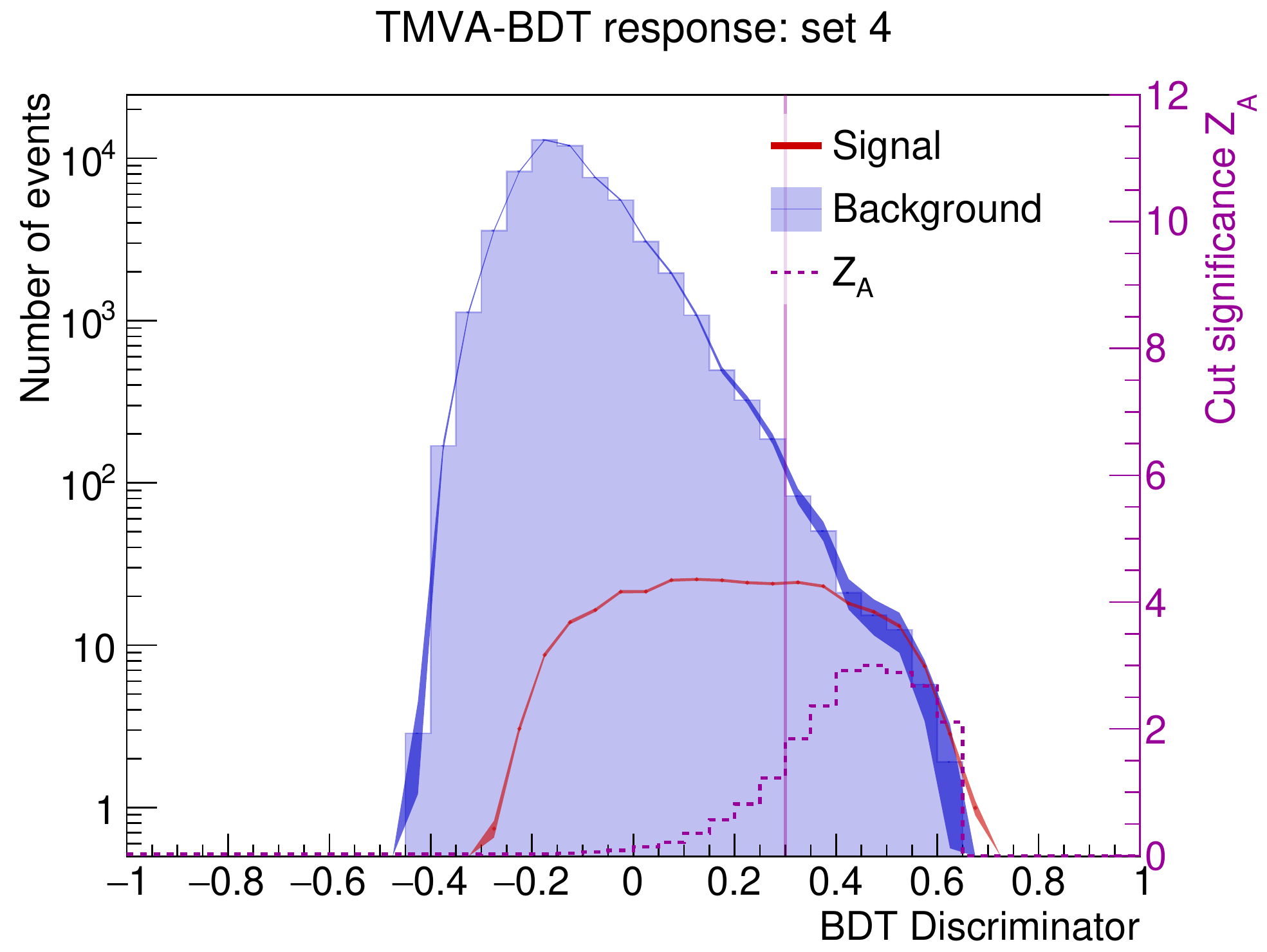}  \label{fig:subtmvabdt}}
\caption{The SVM-HINT and TMVA BDT responses trained with the variables sets set 3 and set 4, as defined in \ref{tab:varsets}. 
Even though the optimal $Z_{A}$ efficiency information is not available in data, it is included to demonstrate the 
reliability of the optimal discriminator cut output from the classifier implementations. The y-axis on the left shows 
the number of events (normalized to the aimed integrated luminosity), whereas the y-axis on the right shows the Asimov 
significance for the discriminator cut per bin.}
\label{fig:hintsvmtmvares2}
\end{figure}

\paragraph{Results}

Figures~\ref{fig:hintsvmtmvares1} and \ref{fig:hintsvmtmvares2} show the performance of the SVM-HINT and the TMVA-BDT
for the four different variable sets. Both the SVM-HINT as well as the TMVA-BDT obtain the highest accuracy with the largest 
number of variables (set 1). Both methods perform much worse with only low-level variables (set 2), while the high-level variables 
(set 3) are clearly able to separate signal and background. Despite the poor performance of the low-level variables, they add
a substantial amount of extra information to enlarge the significance when adding them to the high-level variables. Here we 
conclude that even with variable multiplicity of 25, SVM-HINT as well as TMVA-BDT do not require a preselection of variables. Reducing the number
of variables to two low-level and two high-level variables gives slightly better results for the SVM-HINT than for the TMVA-BDT.
Here we have to note that this study is not meant to compare the two methods, but to put the results obtained by SVM-HINT into a 
more known context in HEP with the TMVA-BDT.
The SVM-HINT with the automated hyperparameter tuning needs no further manual optimization and can make use of high number of variables 
simultaneously with an increasing classification power. 

Furthermore, the importance of the performance measures is visible on the discriminator cut decisions by the classifier 
implementations. The TMVA BDT uses simply $\frac{S}{\sqrt{S+B}}$ as the performance measure, where $S$ and $B$ correspond to the 
number of signal and background events, respectively. This formula performs differently than a log-likelihood significance calculation. 
Therefore, the optimal cut provided by TMVA reduces the significance obtained from the classifier implementation. 
SVM-HINT uses the Asimov significance which gives very similar results to the log-likelihood calculation, and therefore, the results 
obtained from SVM-HINT not only provide good out-of-the-box estimation of the actual significance, but the discriminator cut 
given by SVM-HINT maximizes the significance between background and signal.

% !TEX root = svm_paper.tex
% !TEX spellcheck = en-US
\section{Conclusions}
Our results show that a Support Vector Machine is an efficient machine learning algorithm for new physics searches in high-energy physics. The rare applications of this tool in our field may be related to the limited implementation provided by the popular TMVA library. An appropriate designed automatic search for the two hyper-parameters easily overcomes this limitations and reveals the full potential of this approach. The Support Vector Machine is certainly able to compete with a Boosted Decision Tree which currently is the prevalent machine learning tool in high-energy physics. We do not intent to claim that one of the algorithms out-performs the other. This would need a diligent optimization of our Boosted Decision Tree, which is beyond the scope of this paper. 
The performance of a Boosted Decision Trees depends on a larger number of parameters which complicates the construction of an automated tuning procedure.
The clear advantage of the Support Vector Machine is rather the straightforward hyper-parameter tuning. 
We demonstrate that the approximated median discovery significance (Asimov significance) is an effective figure of merit for the parameter tuning and that only two parameters need to be adapted to define a well performing search tool.
The SVM maximum margin concept guarantees good generalization properties of the trained algorithm while at the same
time the hyper-parameter tuning  allows to find a non-linearly bounded area with maximized significance.
Furthermore, Support Vector Machines are known to be robust against an large number of even partially correlated input variables. This is in agreement with our studies, a lengthy selection of useful input variables was not necessary. 
The algorithm reliably exploits all available information.

\clearpage
% !TEX root = svm_paper.tex

% !TEX root = svm_paper.tex
\appendix
\section{Iterative grid search}\label{app:itergrid} 

\newcommand{\Zt}{\tilde{Z}_A}
\newcommand{\Ztl}{\tilde{Z}^{(l)}_A}
As shown in section~\ref{sec:hyper} an SVM with RBF kernel requires two different hyper-parameters to be adjusted: $C$ and $\gamma$. While in principle a \textsl{brute force} grid search is sufficient to find the best hyper-parameters, we used an adaptive search strategy for the results in this paper which we describe here for completeness. 
%An SVM problem with the RBF kernel requires two different hyper-parameters to be adjusted: $C$ and $\gamma$. With the increasing $\gamma$, the kernel becomes more sensitive to the non-linear shape of the distributions and more vulnerable to overtraining. Therefore, the optimum $\gamma$ value depends on the variables' topology. The $C$ value adjusts the impact of slack variables in the minimization problem. Optimization of these parameters plays a critical role for the performance of the SVM classifier.\\
The SVM-HINT grid search algorithm uses a modified version of the Asimov Significance \ref{asimovz}, a significance score $\Zt$ based on the difference between the significance value observed in the test sample and the significance value from the training sample. 
%\begin{equation}
%\Zt = 1-|Z_A^{(test)}-Z_A^{(train)}|/(Z_A^{(test)}+Z_A^{(train)}).
%\end{equation}
\begin{equation}
\Zt = Z_A^{(test)}\left[1-\frac{|Z_A^{(test)}-Z_A^{(train)}|}{Z_A^{(test)}+Z_A^{(train)}}\right].
\end{equation}
This way, the extreme significance values observed due to fluctuations or overtraining can be penalized without a high computational effort. 
The search algorithm %introduced in the SVM HEP interface (SVM-HINT) 
can be formalized as follows:
\begin{enumerate}
\item {For the given initial parameters $\gamma_{initial}$ and $C_{initial}$, the iterative grid search algorithm produces an array of logarithmically spaced $\gamma$ values with a step size $K_t$ around the mid-value $\gamma^{(1)}_m=\gamma_{initial}$ such that:
\begin{equation}
\begin{aligned}
\gamma^{(l)}_k =&\; K_t\cdot \gamma^{(l)}_{k-1},\quad \text{where}\quad K_t=\frac{1}{2}(1+\ln(t/2)),\\ 
k=&\; 0,\dots ,m,\dots,2m=18, \qquad t=\integer(l/4),\qquad l=1,\dots,20 %\lfloor\frac{l}{4}\rfloor
\end{aligned} \label{eqn:gamma}
\end{equation}
where $l$ indicates the number of iterations, the variable $t$ is a \textsl{focus parameter} that decreases the step size factor $K_t$  every fourth iteration. \\$\Zt$ is then evaluated for all of these $C$-$\gamma$-pairs. 
%every four iterations% and $e$ is the natural logarithm constant. One thing to note here is that $\gamma_{initial}$ becomes the median of the gamma array.  
}
\item For the next step $C$ is increased to $C^{(l+1)}=1.5\cdot C^{(l)}$ and $\Zt$ 
%For each $\gamma_k$, $Z^{(l)}_A$ is calculated with the initial $C_{initial}$ and the initial $C$ value increased in the next iteration by $0.5 \times C$, and $Z^{(l)}_A$ 
is again calculated with each value in the $\gamma$ array. 
\item If the maximum $\Ztl$ value is at least $30\,$\% larger then the best $\Zt^{l-1}$ from the previous iteration the higher $C$ parameter is accepted. The $30\,$\% hurdle is introduced to stabilize against fluctuations.
%If the maximum $Z^{(l)}_A$ obtained from the various $\gamma$ values in the gamma array of the iteration $l$ is less than the %sixty-eight percent of the $l-1$, the $\gamma$ and $C$ values of iteration $l-1$'s maximum $Z^{(l-1)}_A$ are taken as the %optimal hyper-parameters. 
\item After each fourth iteration, the $C$-$\gamma$-pair corresponding the highest significance score is taken as the new initial $\gamma$ and the algorithm returns to the first step; now with a smaller step size factor $K_t$
%the increased \textsl{focus parameter}. Therefore, 
such that the new $\gamma$ array has a tighter stepping around the new initial $\gamma$ value.% than the previous $\gamma$ array. 
\item When the number of iterations reaches the pre-defined maximum value, the algorithm stops and the $\gamma$-$C$-pair with the maximum $\Zt$ in the final iteration are returned as the best hyper-parameter values. 
\end{enumerate}
The procedure assumes that a sufficiently small $C_{initial}$ had been chosen. In case that the found best $C$ value is identical with the $C_{initial}$ the algorithm is restarted with a smaller value of $C_{initial}$.

% !TEX root = svm_paper.tex
% !TEX spellcheck = en-US
\section{Boosted Decision Trees}\label{BDT}
%DK: should we compare to a BDT with Asimov as FOM?
%We must mention what is used later
\begin{figure}[h] 
\centering
\includegraphics[width=0.4\textwidth]{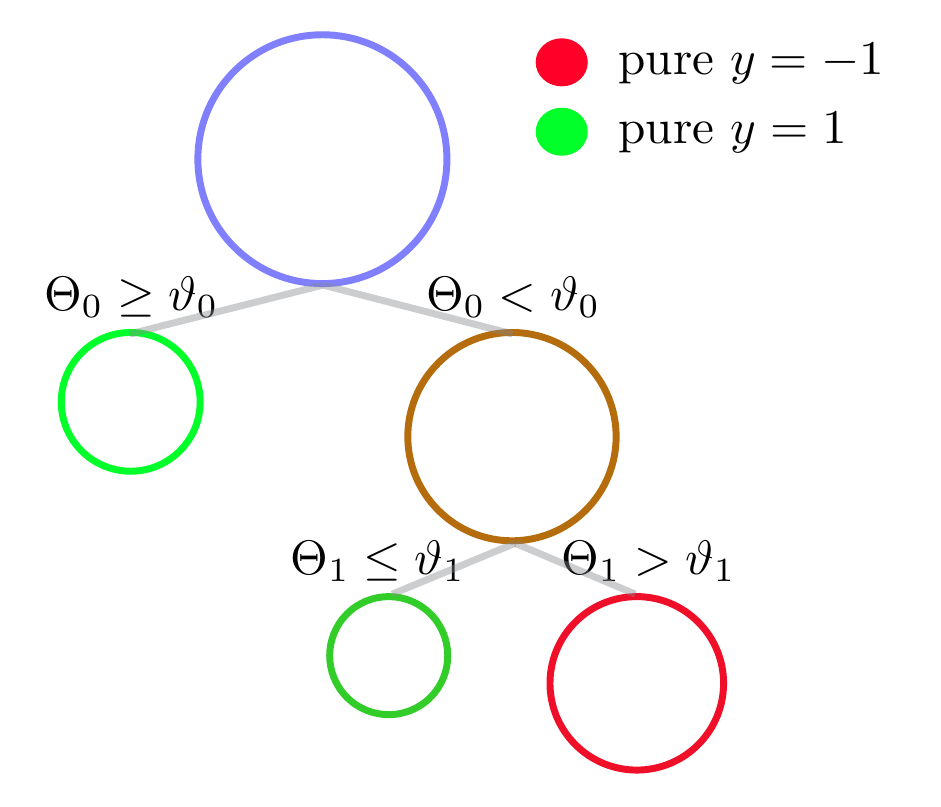}
\caption{Representation of a simple binary decision tree structure: Each node is split with respect to a variable $\Theta_i$ and a cut value $\theta_i$ determined by the performance measure.}
\label{fig:bdt}
\end{figure}
Boosted decision trees (BDT) are probably the most common ML classifier in experimental particle physics. We therefore compare the performance of our SVM framework to a BDT implemented with TMVA. We shortly introduce the relevant BDT concepts used in this comparison.

\paragraph{Decision Trees}\label{DT}
A binary decision tree \cite{CART84} separates signal and background by a sequence of binary cuts (Fig.~\ref{fig:bdt}).
The terminating branches, or leaves, correspond to a cubical separation in the multi-dimensional space of the input variables and the tree as a whole forms a complex separation boundary between the two classes.
Each node of the tree is connected to two branches that are split with respect to only one of the variables  $x_i^{(1)}\dots x_i^{(n)}$ defined in \ref{for:tra2}.
At each node the algorithm selects one of the available physical variables and searches for a best cut value. 
%Even for relatively small sized trees simultaneously calculating all possible combinations and cut values would be a lengthy procedure - actually it is NP-hard. 
%Starting with the root node the optimization is typically done node by node. 
This process requires a suitable goodness-of-split measure % but in practice the final results do not strongly depend on the particular choice~\cite{Hocker:2007ht}. 
and a common choice is the \textsl{Gini impurity} index which is also used for the TMVA-BDT in this paper.
For the two class case the \textsl{Gini} index is given by 
$
g = 2 p (1-p),
$
where the purity $p$ of a node is defined as the ratio of signal events over all events. 
%(the definition implies that a pure background node has purity zero) 
%Starting with the root node 
The training starts with the root node, and the tree is constructed recursively while at each split the reduction in \textsl{Gini} impurity is maximized. The splitting stops when a node falls below a predefined minimum of events.% (TMVA: \verb#nEventsMin#).  
\paragraph{Pruning}
The constructed decision tree is sensitive to statistical fluctuation. To avoid overtraining it must be pruned to remove statistically insignificant nodes.
\textsl{Cost complexity pruning} \cite{CART84} removes branches which increase the misclassification cost.  The misclassification rate 
$R=1-\max(p,1-p)$ is used as a cost estimate at each node and compared to the cost of the subtree below the node. The cost complexity is defined as 
$
\rho=(R_{\mbox{node}}-R_{\mbox{subtree}})/(N_{st} - 1),
$
where $N_{st}$ is the number of nodes in the subtree.
The node with the smallest $\rho$ is recursively removed from the tree as long as $\rho$ is below a certain pruning strength value $\rho_0$.
\paragraph{Boosting}
A single decision tree is seldom an efficient classifier.  Boosting is a powerful iterative algorithm which improves the performance of  weak classifiers. 
The boosting of a decision tree extends this concept from one tree to a forrest of trees. The trees are derived from the same training data by reweighting events, and are finally combined into a single classifier of considerably enhanced performance. For this paper AdaBoost \cite{AdaBoost97} is used. %\cite{Schapire:2012aa}.

The TMVA-BDT (AdaBoost) is trained and tested with 32 different settings to obtain an optimal parameter set 
(TMVA: \verb#nEventsMin#, \verb#NTrees#, \verb#MinNodeSize#, \verb#MaxDepth# and \verb#AdaBoostBeta#) and the best performing configuration is used for evaluation\footnote{Configuration files are available at \cite{ourgithub}}.

% TMVA first boosting then pruning
%!!! For this part out main reference should definitely be Classification and Regression Trees, Breiman, Friedman, Olshen, Stone and for the Boosting (adaboost) Freund and Schapire's paper !!!  

%\input{Implementations}

\acknowledgments
M. \"O. Sahin would like to thank the Joachim Herz foundation for the support. 

% query inspire, trying to verify the data (this will probably spot
\bibliographystyle{utphys}
\bibliography{lit}
%\begin{thebibliography}{99}
%\bibitem{a}
%Author, \emph{Title}, \emph{J. Abbrev.} {\bf vol} (year) pg.
%\bibitem{b}
%Author, \emph{Title},
%arxiv:1234.5678.
%\bibitem{c}
%Author, \emph{Title},
%Publisher (year).
% Please avoid comments such as "For a review'', "For some examples",
% "and references therein" or move them in the text. In general,
% please leave only references in the bibliography and move all
% accessory text in footnotes.
% Also, please have only one work for each \bibitem.
%\end{thebibliography}
\end{document}